\documentclass[bibyear,traditabstract]{aa} 
\usepackage{epsfig}
\usepackage{threeparttable}
\usepackage{multicol}
\usepackage{graphicx,times,multirow,amsmath,amssymb,color,longtable,breqn}
\usepackage[T1]{fontenc}
\usepackage[varg]{txfonts}
\usepackage{booktabs}

\def\ltsima{$\; \buildrel < \over \sim \;$}
\def\lsim{\lower.5ex\hbox{\ltsima}}
\def\gtsima{$\; \buildrel $\geq$ \over \sim \;$}
\def\gsim{\lower.5ex\hbox{\gtsima}}
\newcommand{\be}{\begin{equation}}
\newcommand{\en}{\end{equation}}

\def\deg {$^\circ$}
\def\nh{\hbox{$N_{\rm H}$}}
\def\flux {\mbox{erg~cm$^{-2}$~s$^{-1}$}}
\def\lum {\mbox{erg~s$^{-1}$}}
\def\src{J1109}

\newcommand{\cxo}{{\em Chandra}}
\newcommand{\xmm}{{\em XMM--Newton}}
\newcommand{\swift}{{\em Swift}}
\newcommand{\fermi}{{\em Fermi}}
\newcommand{\nustar}{{\em NuSTAR}}
\newcommand{\integ}{INTEGRAL}
\newcommand{\ros}{ROSAT}
\newcommand{\psr}{PSR\,J1023$+$0038} 
\newcommand{\xss}{XSS\,J12270$-$4859}
\newcommand{\parallax}{\ensuremath{\varpi}}

\begin{document}

\title{Prolonged sub-luminous state of the new transitional pulsar candidate CXOU\,J110926.4$-$650224}
\author{Francesco Coti Zelati\inst{1,2}, Alessandro Papitto\inst{3}, Domitilla de Martino\inst{4}, David A. H. Buckley\inst{5}, 
Alida Odendaal\inst{6}, \\ Jian Li\inst{7}, Thomas D. Russell\inst{8}, Diego F. Torres\inst{1,2,9}, Simona M. Mazzola\inst{10}, 
Enrico Bozzo\inst{11}, Mariusz Gromadzki\inst{12}, \\ Sergio Campana\inst{13}, Nanda Rea\inst{1,2}, Carlo Ferrigno\inst{11}, 
and Simone Migliari\inst{14,15}} 
\institute{Institute of Space Sciences (ICE, CSIC), Campus UAB, Carrer de Can Magrans s/n, 08193 Barcelona, Spain\\
 \email{cotizelati@ice.csic.es}
 \and
 Institut d'Estudis Espacials de Catalunya (IEEC), Gran Capit\`a 2-4, 08034 Barcelona, Spain
 \and
 INAF -- Osservatorio Astronomico di Roma, via Frascati 33, 00040 Monte Porzio Catone (Roma), Italy
 \and
 INAF -- Osservatorio Astronomico di Capodimonte, Salita Moiariello 16, 80131 Napoli, Italy
 \and
 Southern African Large Telescope Foundation, PO Box 9, 7935 Observatory, Cape Town, South Africa
 \and
 Department of Physics, University of the Free State, PO Box 339, 9300 Bloemfontein, South Africa
  \and
 Deutsches Elektronen-Synchrotron (DESY), 15738 Zeuthen, Germany
  \and 
 Anton Pannekoek Institute for Astronomy, University of Amsterdam, Science Park 904, 1098 XH Amsterdam, The Netherlands
 \and
 Instituci\'o Catalana de Recerca i Estudis Avan\c cats (ICREA), 08010 Barcelona, Spain
 \and
 Dipartimento di Fisica e Chimica, Universit\`a di Palermo, via Archirafi 36, 90123 Palermo, Italy
 \and
 Department of Astronomy, University of Geneva, Chemin d'Ecogia 16, 1290 Versoix, Switzerland
 \and
 Warsaw University Astronomical Observatory, Al. Ujazdowskie 4, 00-478 Warszawa, Poland
 \and
 INAF -- Osservatorio Astronomico di Brera, via Emilio Bianchi 46, 23807 Merate (LC), Italy
 \and
 XMM-Newton Science Operations Centre, ESAC/ESA, Camino Bajo del Castillo s/n, Urb. Villafranca del Castillo, \\28691 Villanueva de la Ca\~nada, Madrid, Spain
 \and
 Department of Quantum Physics and Astrophysics \& Institute of Cosmos Sciences, University of Barcelona, Mart\'i i Franqu\`es 1, 08028 Barcelona, Spain
 }

\date{Received 12 December 2018  / Accepted 26 December 2018}

\abstract{We report on a multi-wavelength study of the unclassified X-ray source CXOU\,J110926.4$-$650224 (\src). We identified the optical counterpart as a blue star with a magnitude of $\sim20.1$ (3300--10\,500~\AA). The optical emission was variable on timescales from hundreds to thousands of seconds. The spectrum showed prominent emission lines with variable profiles at different epochs. Simultaneous \xmm\ and \nustar\ observations revealed a bimodal distribution of the X-ray count rates on timescales as short as tens of seconds, as well as sporadic flaring activity. The average broad-band (0.3--79~keV) spectrum was adequately described by an absorbed power law model with photon index of $\Gamma=1.63\pm0.01$ (at 1$\sigma$ c.l.), and the X-ray luminosity was $(2.16\pm0.04)\times10^{34}$~\lum\ for a distance of 4~kpc. Based on observations with different instruments, the X-ray luminosity has remained relatively steady over the past $\sim 15$~years. \src\ is spatially associated with the gamma-ray source FL8Y\,J1109.8$-$6500, which was detected with \fermi\ at an average luminosity of $(1.5\pm0.2)\times10^{34}$~\lum\ (assuming the distance of \src) over the 0.1--300~GeV energy band between 2008 and 2016. The source was undetected during ATCA radio observations that were simultaneous with \nustar, down to a 3$\sigma$ flux upper limit of 18~$\mu$Jy/beam (at 7.25~GHz). We show that the phenomenological properties of \src\ point to a binary transitional pulsar candidate currently in a sub-luminous accretion disk state, and that the upper limits derived for the radio emission are consistent with the expected radio luminosity for accreting neutron stars at similar X-ray luminosities.}


\titlerunning{The transitional pulsar candidate CXOU\,J110926.4$-$650224}
\authorrunning{F. Coti Zelati et al.} 

\maketitle

\section{Introduction}
\label{sec:introduction}

Radio millisecond pulsars (MSPs) are believed to be old, weakly magnetised pulsars that have been spun up after a Gyr-long phase of angular momentum transfer through mass accretion from a sub-solar ($M\lesssim1M_{\sun}$) companion star in a low-mass X-ray binary system (LMXB; e.g. Alpar et al. 1982). Swings between a rotation-powered MSP state and an accretion-powered LMXB state were observed recently over a few weeks in the transient system \object{IGR\,J18245$-$2452} (Papitto et al. 2013), proving that the evolutionary path towards a radio MSP is not smooth, but encompasses transitions on timescales compatible with changes of the mass accretion rate onto the neutron star (NS). State transitions have also been observed recently in the systems \object{\psr} (Archibald et al. 2009; Stappers et al. 2014) and \object{\xss} (Bassa et al. 2014), although these have never been observed to undergo a fully-fledged accretion outburst (see Campana \& Di Salvo 2018, for a review). 

The three `transitional' MSPs (tMSPs) share an enigmatic intermediate sub-luminous accretion disk state where the X-ray luminosity is orders of magnitude lower than that measured during the typical accretion outbursts of NS LMXBs. The X-ray emission shows a peculiar bimodal distribution in the count rates over a broad energy range (up to at least 80~keV; e.g. Tendulkar et al. 2014). The switches between the two different intensity levels (henceforth dubbed `modes') occur on timescales as short as tens of seconds. Sporadic flaring activity is also observed (e.g. de Martino et al. 2013; Coti Zelati et al. 2018). tMSPs in the sub-luminous accretion disk state are observed to emit in the gamma-ray band, with a luminosity comparable to that in the X-ray band (e.g. Torres et al. 2017), as well as in the radio band. In particular, the radio emission is characterised by a flat to slightly inverted spectrum, and is rapidly variable, a factor of approximately two to three on timescales of minutes (e.g. Deller et al. 2015). Coherent X-ray pulsations at the NS spin period have been observed only in the high mode, and were initially interpreted in terms of intermittent mass accretion onto the NS surface (Archibald et al. 2015; Papitto et al. 2015). However, the radio flaring activity associated with the X-ray low mode (Bogdanov et al. 2018) and the detection of optical pulses during the X-ray high and flare modes (Ambrosino et al. 2017; Papitto et al. submitted) in the prototypical tMSP, \psr, challenge this interpretation, and suggest instead an active rotation-powered MSP in the sub-luminous accretion disk state. 

\begin{figure}
\begin{center}
\includegraphics[width=.59\textwidth]{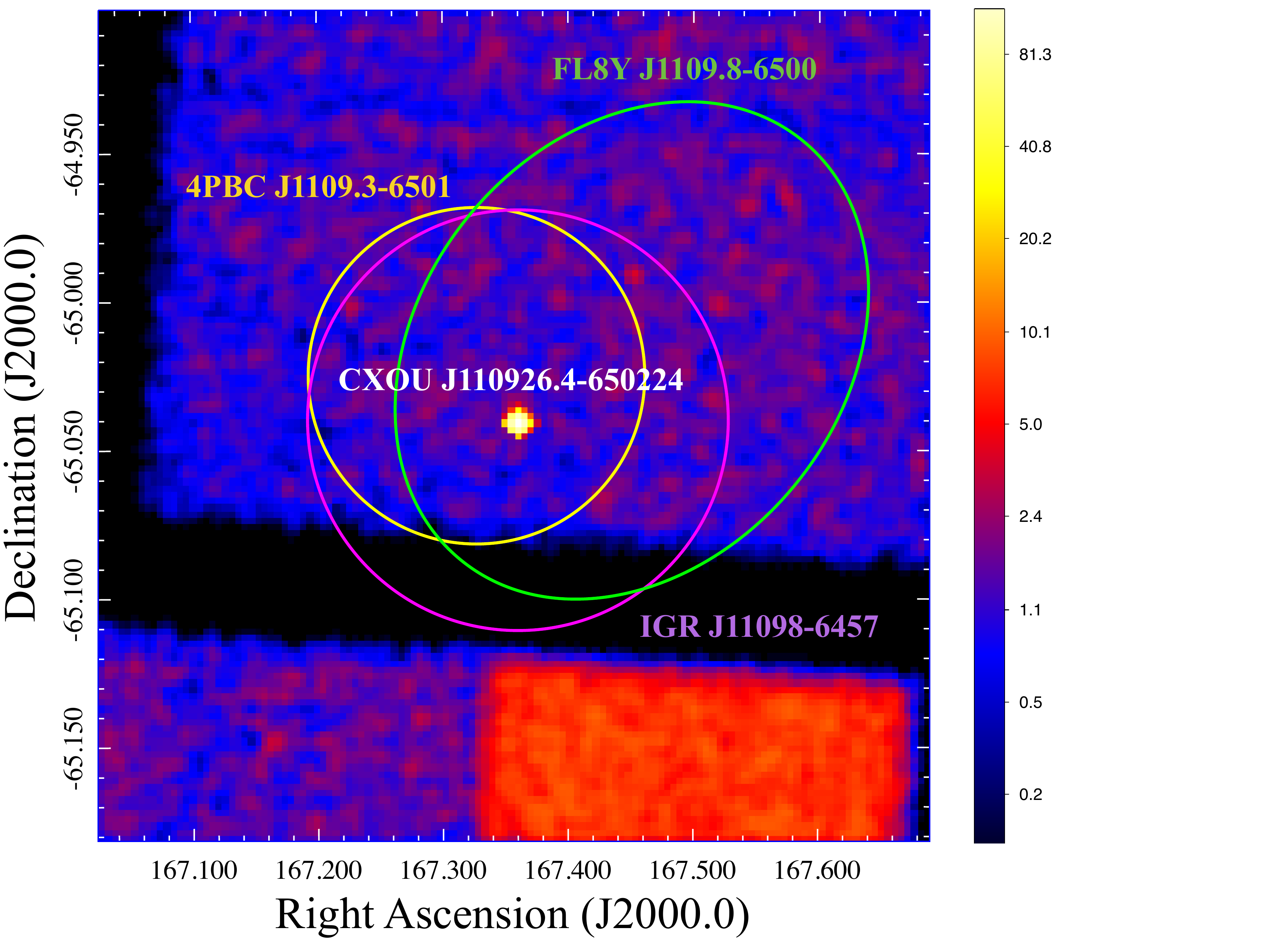}
\caption{Exposure-corrected image of the field around \src\ (labelled in white) extracted from \cxo\ ACIS-I data over the 0.3--8~keV energy band. The image was smoothed with a Gaussian filter with a kernel radius of 3 pixels (one ACIS pixel corresponds to 0.492 arcsec), and a logarithmic scale was adopted for better visualization. North is up, east to the right. The green ellipse has semi-major and semi-minor axes of 5.5 and 4.3 arcmin, respectively, at a position angle of 39.2\deg\ (measured from north towards east), and represents the error region (at 95\% c.l.) for the position of the gamma-ray source FL8Y\,J1109.8$-$6500 obtained using 8 years of \fermi/LAT data. The magenta circle represents the error region for the position of the hard X-ray source \object{IGR\,J11098$-$6457}, obtained using 2995~ks of \integ\ IBIS data between 2002 and 2010 (see the 8~yr \integ\ IBIS soft gamma-ray survey catalogue by Bird et al. 2016 for more details). The yellow circle represents the error region for the position of the hard X-ray source 4PBC\,J1109.3$-$6501, obtained using 32.8~Ms of \swift\ BAT data between 2004 December and 2013 March (see the 100-month Palermo \swift\ BAT catalogue by Cusumano et al. 2014 for more details). The colour code on the right indicates the S/N at the different positions in the image.}
\label{fig:fov}
\end{center}
\end{figure}

In search for X-ray counterparts to the 2131 unclassified gamma-ray sources in the recent preliminary release of the \fermi\ Large Area Telescope (LAT) eight-year point source list\footnote{This list is preliminary as reported at \texttt{https://fermi.gsfc.nasa.gov/ssc/data/access/lat/fl8y/}}, we pinpointed CXOU\,J110926.4$-$650224 (\src) as the only X-ray counterpart to FL8Y\,J1109.8$-$6500. \src\ was discovered during a $\sim5$-ks observation with \cxo\ in 2008, and singled out as the soft X-ray counterpart of the unidentified persistent hard X-ray source \object{IGR\,J11098$-$6457} (Tomsick et al. 2009). It is also listed in the Palermo \emph{Neil Gehrels Swift Observatory} (\swift) Burst Alert Telescope (BAT) source catalogue as 4PBC\,J1109.3$-$6501\footnote{See \texttt{http://bat.ifc.inaf.it/100m_bat_catalog/100m_\\bat_catalog_v0.0.htm}} (Cusumano et al. 2014)(see Fig.~\ref{fig:fov}). However, its nature has so far remained elusive. The spatial coincidence with the hard X-/gamma-ray source, as well as the dramatic time variability of the soft X-ray emission on timescales of tens of seconds that we revealed from a preliminary analysis of the \cxo\ data (see Sect.~\ref{sec:xrayresults} for more details), represent phenomenological properties shared by all three tMSPs in the sub-luminous accretion disk state. We hence suspected that \src\ could be a tMSP in the same state.

\begin{figure}
\begin{center}
\includegraphics[width=.485\textwidth]{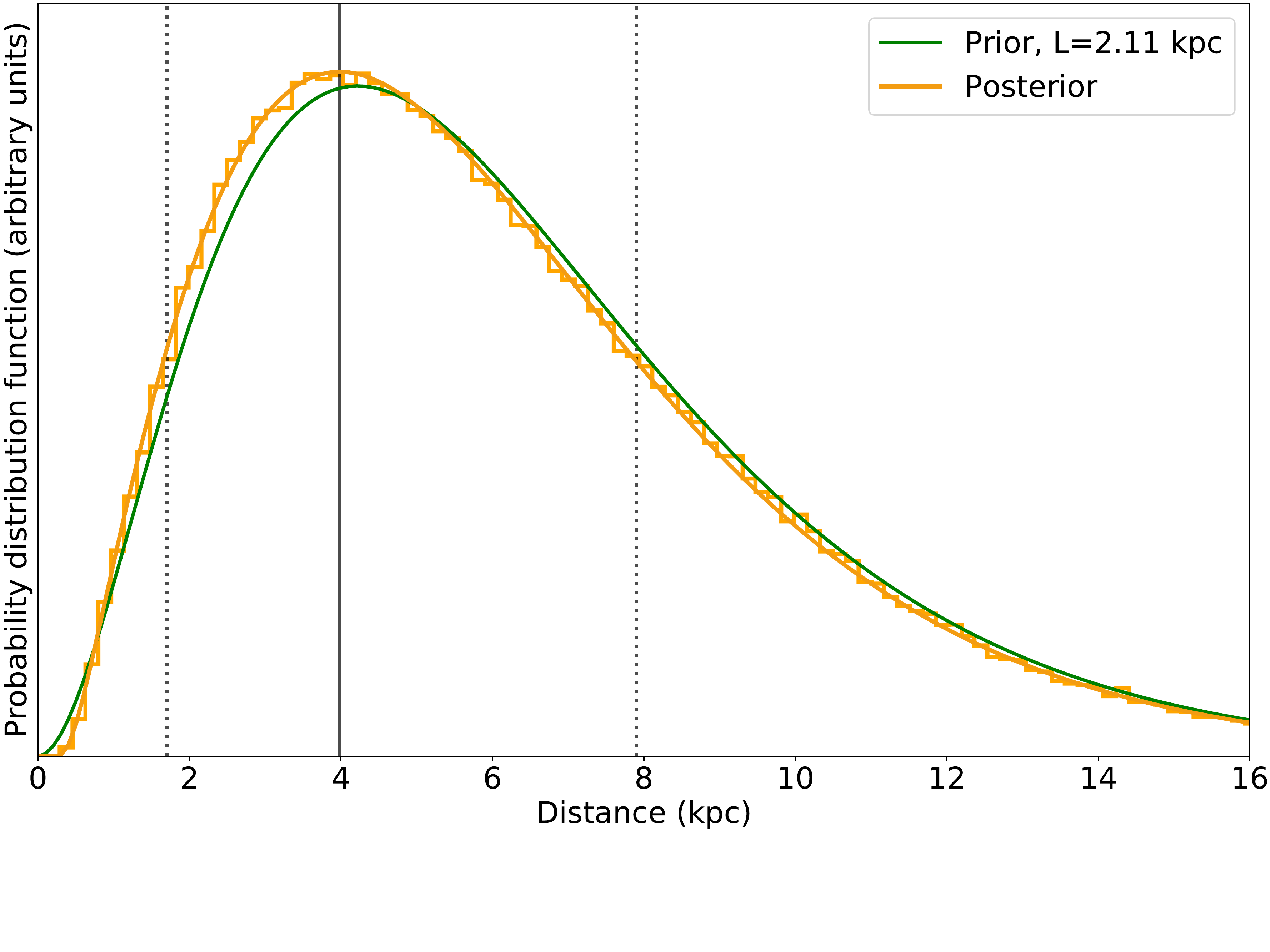}
\vspace{-1cm}
\caption{Posterior probability distribution function (PDF; yellow curve) for the distance of \src\ for an exponentially decreasing volume density prior with length scale of $L=2.11$~kpc (green curve). The mode of the PDF is indicated by the vertical black solid line. The lower and upper boundaries of the highest density interval containing 68.27\% of the posterior probability (i.e., the 16th and 84th percentiles) are indicated by the two vertical black dotted lines. The known systematic zero-point offset of -29~$\mu$as in the {\it Gaia} parallaxes determined from {\it Gaia} observations of quasars (Lindegren et al. 2018) was taken into account in the computation. The plot was produced using the code publicly available at \texttt{https://github.com/Alymantara/gaia_distance_calculator/} \texttt{blob/master/gaia_distance_calculator.ipynb}.}
\label{fig:distance}
\end{center}
\end{figure}

This paper reports on a multi-wavelength study of \src\ aimed at ascertaining its nature. The manuscript is structured as follows: we report details on the associated optical and gamma-ray sources in Sect.~\ref{sec:assoc}; we present the optical, X-ray, and radio observations and the corresponding data analysis in Sects.~\ref{sec:optical}--\ref{sec:radio}, respectively; we report on the results in Sect.~\ref{results}; discussion of our results and conclusions follow in Sects.~\ref{discuss} and ~\ref{conclusions}, respectively.

\section{The optical and gamma-ray counterparts}
\label{sec:assoc}

The X-ray source is located at $\rm RA=11^h09^m26^s.43$, $\rm Dec= -65^\circ02'25''.0$ (J2000.0 equinox), with a positional uncertainty of 0.55 arcsec at a 1$\sigma$ confidence level (c.l.)\footnote{The uncertainty radius was computed as the sum of the positional systematic uncertainty of the ACIS (Weisskopf 2005) and the statistical uncertainty derived from the source counts and its distance from the aim point (see Eq.~(14) by Kim et al. 2007).}. The error circle includes an optical source listed in the second {\it Gaia} data release (Gaia Collaboration 2018) at $\rm RA=11^h09^m26\fs4057(9)$, $\rm Dec= -65^\circ02'24\farcs8216(4)$ (J2015.5 Barycentric Coordinate Time; 1$\sigma$ uncertainties on the last digits are reported in parentheses). Its average observed magnitudes are $G\sim20.1$, $B_p\sim20.2$ and $R_p\sim19.2$ over the 3300--10\,500~\AA, 3300--6800~\AA\ and 6400--10\,500~\AA\ wavelength ranges, respectively\footnote{$B_{{\rm p}}$ and $R_{{\rm p}}$ represent the magnitudes measured by the Blue and Red Photometers on board {\it Gaia}, respectively.}. Its absolute trigonometric parallax, $\parallax=0.73 \pm 1.06$~mas, can be converted into a distance by means of a Bayesian method, assuming an exponentially decreasing volume density prior of the form $P(D~|~L) = (2L^3)^{-1} D^2 {\rm~exp}(-D/L)$ (see Bailer-Jones et al. 2018). For such an uncertain measurement, the distance estimate is dominated by the choice of the prior length scale. On the one hand, a value of $D =4.0$~kpc with 16th and 84th percentiles of 1.7 and 7.9~kpc, respectively, is obtained when adopting a length scale of $L=2.1$~kpc, based on a three-dimensional model of the Galaxy as seen by {\it Gaia} (Rybizki et al. 2018). On the other hand, a larger value of $D =6.5$~kpc with 16th and 84th percentiles of 4.4 and 15.5~kpc, respectively, is derived when considering a length scale of $L=3.4$~kpc, based on the three-dimensional model of the space distribution of LMXBs detected in the X-ray band in the Galaxy (Grimm et al. 2002; Gandhi et al. 2018). The posterior probability distribution function for the former case is shown in Fig.~\ref{fig:distance}.

The gamma-ray source was detected with the \fermi/LAT (Atwood et al. 2009) at an energy flux of $F_{\gamma} = (6\pm1) \times 10^{-12}$ \flux\ over the 0.1--100~GeV energy band between 2008 August and 2016 August. Its spectrum was well described by a power law model with photon index of $\Gamma_{\gamma} = 2.58\pm0.15$.
No information at higher energy is available, as the region around \src\ was not covered by the {\it H.E.S.S.} survey of the Galactic plane (H.E.S.S. Collaboration 2018{\footnote{See \texttt{https://www.mpi-hd.mpg.de/hfm/HESS/hgps/}}}).

In the following, we will adopt the position of the optical counterpart as a reference for all datasets, assume for simplicity a distance of $D=4$~kpc, and quote all uncertainties at the 1$\sigma$ c.l. (unless otherwise explicitly stated). The gamma-ray flux measured with \fermi/LAT translates into a luminosity of $L_\gamma =(1.5\pm0.2)\times10^{34}~d_4^2$ \lum\ over the 100 MeV--300~GeV energy band, where $d_4$ is the distance in units of 4~kpc. We stress, however, that the assumed value for the distance should be taken with caution, owing to the very large uncertainty on the parallax measurement.

\section{Optical observations}
\label{sec:optical}

\subsection{\textit{SALT}}
\label{sec:salt}

We obtained a set of six consecutive 100-s images of the field around \src\ in the $g'$, $r'$ and $i'$ SDSS filters with the SALTICAM imaging camera (O'Donoghue et al. 2006) mounted on the 10 m Southern African Large Telescope (SALT; Buckley et al. 2006) at the Sutherland plateau of the South African Astronomical Observatory. The observations started on 2018 March 4 at 00:52:29.33 Coordinated Universal Time (UTC). 

We also obtained optical spectra of \src\ using the Robert Stobie Spectrograph (RSS) on SALT in long-slit mode (Burgh et al. 2003; Kobulnicky et al. 2003).  Two spectra each with exposure times of 1000~s were recorded on 2018 March 4 (starting at 01:25:05 and 01:41:58 UTC, respectively), and one with exposure time of 1800~s on 2018 July 6 (starting at 17:19:17 UTC). The PG0300 grating was used at grating angles of 5.375\deg\ and 5\deg\ respectively for the two dates, yielding a usable wavelength range of $\sim$3700-9000~\AA\ and a resolving power of 350 at 6200~\AA. We used a 1.5-arcsec slit placed at a position angle of ${\rm PA} = 20$\deg\ (measured from north towards east) for all the observations.

The reductions of the photometric and spectroscopic data were performed with \texttt{PySALT} version 0.47, the official \texttt{PyRAF}-based software package for SALT data reductions (Crawford et al. 2010)\footnote{\texttt{https://astronomers.salt.ac.za/software/pysalt-\\documentation/}}, and included gain correction, correction for cross-talk between the amplifiers, bias subtraction and amplifier mosaicking. Absolute flux calibration of SALT data is not feasible, owing to the moving pupil of the telescope (the effective area of the telescope constantly changes during the tracks and exposures; see Buckley et al. 2008).

\begin{figure*}
\centering
\includegraphics[width=1.04\textwidth]{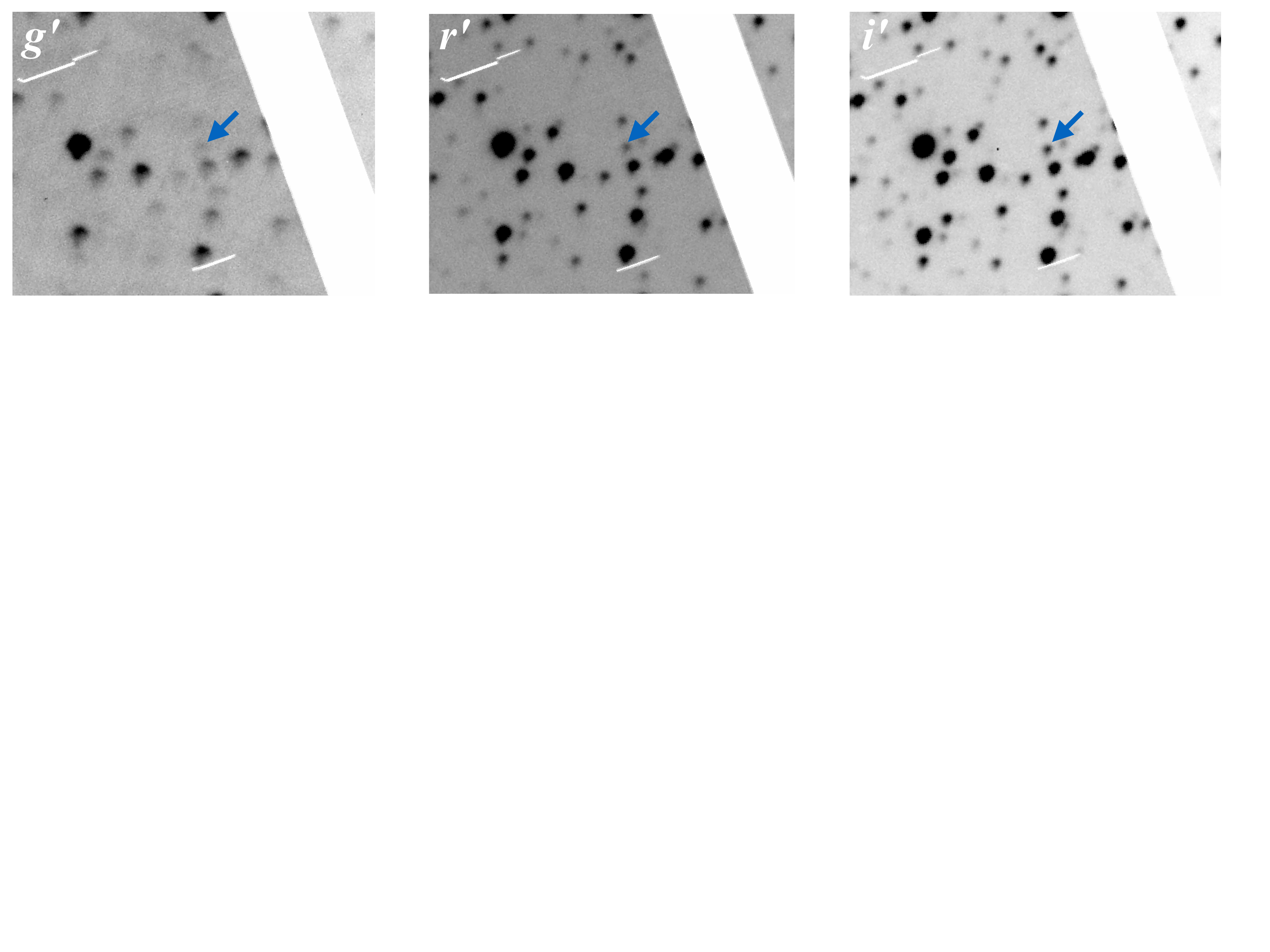}
\vspace{-10cm}
\caption{Zoom of the optical images of the field around \src\ obtained on 2018 March 4 using SALTICAM. North is up, east to the left. Images acquired in the same filter were co-added, yielding total net exposures of 300~s, 200~s and 100 s in the $g'$, $r'$ and $i'$ filters, respectively. The white strip is due to the gap between the two CCD chips. The blue arrowhead in each panel indicates the position of the source in the second {\it Gaia} data release located within the error circle of \src\ (as determined using \cxo\ data). See the text for details.}
\label{fig:salticam_images}
\end{figure*}

The cleaned images of the field around \src\ in the three filters are shown in Fig.~\ref{fig:salticam_images}. 
We performed differential photometry of \src\ against three relatively bright close-by (<30 arcsec) stars listed in the Two Micron All Sky Survey (2MASS; Skrutskie et al. 2006) all-sky point source catalogue, using the \texttt{daophot} task (Stetson 2000) within the \texttt{IRAF} (Image Reduction and Analysis Facility) package\footnote{\texttt{http://iraf.noao.edu/}} (Tody 1986, 1993). We derived the $g'$, $r'$ and $i'$ magnitudes of the three stars using their $G$, $B_p$ and $R_p$-band magnitudes quoted in the second {\it Gaia} data release and the magnitude colour conversion formulas reported by Evans et al. (2018; see their Table~A.2), after having verified consistency in the conversion between the 2MASS and {\it Gaia} magnitudes. This approach allowed us to minimise any systematic effect related to flat-fielding, vignetting and atmospheric turbulence. For each acquired image, we then averaged the values for the magnitudes obtained using the three stars. 

We performed further processing on the spectroscopic data through standard spectral calibration procedures with \texttt{IRAF}. Wavelength calibration was performed using an Argon arc lamp spectrum acquired immediately after the science exposures. Background subtraction was done by fitting and subtracting a cubic spline function along every constant-wavelength column in the two-dimensional spectral image, fitting only the background and ignoring stellar spectra and cosmic rays in the image.  During the spectral extraction from a tilted rectangular aperture, a cleaning algorithm was applied to remove cosmic rays from the resulting spectrum. Relative flux calibration was performed for all the spectra, using observations of spectrophotometric standard stars at exactly the same RSS configurations as the \src\ observations.  The 2018 March 4 target spectra were calibrated using a spectrum of \object{HILT600} obtained on 2018 February 28, while the 2018 July 6 target spectrum was calibrated with a spectrum of \object{LTT1020} obtained on 2018 November 11.

\subsection{XMM--Newton optical monitor}

The Optical/UV Monitor Telescope (OM; Mason et al. 2001) on board \xmm\ (Jansen et al. 2001) observed \src\ starting on 2018 June 20 at 20:05:09 UTC (observation ID: 0764344301; PI: A. Papitto). It collected the data both in the `Image' and `Fast Window' modes using the $B$ filter (3975--5025~\AA). Only seven exposures were acquired in the image mode (each lasting 4.3--4.5~ks), owing to an instrument anomaly during the observation (Ballo, priv. commun.). 

We reduced the data using the \texttt{omichain} and \texttt{omfchain} processing pipelines of the \xmm\ Science Analysis System (\texttt{SAS} v.~17.0.0). 
We corrected residual boresight errors in each image by cross-matching the positions of the sources detected with those appearing in the second {\it Gaia} data release.

\section{X-ray observations}
\label{sec:xrays}

\begin{table*}
\small
\begin{center}
\caption{Journal of all X-ray pointed observations of \src.} 
\label{tab:log}
\resizebox{2.04\columnwidth}{!}{
\begin{tabular}{cccccccc} \hline \hline 
Satellite 	    	& Obs. ID			& Instrument 		& Mode	 	&  Start 	-- End time  						& Net exposure & Net count rate 				\\
         	    	&          			&       			&			& (UTC)			     						& (ks)   		& (counts s$^{-1}$)	\\
\hline \vspace{0.15cm}			
\swift\ 		& 00037050001	& XRT			& PC  	 	& 2008 Jan 22 00:18:35 --  2008 Jan 22 11:33:56 	& 3.4  		& $0.041\pm0.003$		\\ \vspace{0.15cm}
\swift\ 		& 00037050002	& XRT			& PC  		& 2008 Feb 11 18:50:37 --  2008 Feb 11 19:02:57 	& 0.7			& $0.025\pm0.006$		\\ \vspace{0.15cm}
\cxo\  		& 9066			& ACIS-I			& TE FAINT  	& 2008 Sep 12 14:39:22  -- 2008 Sep 12 16:44:12 	& 5.1  		& $0.182\pm0.006$		\\ \vspace{0.15cm}
\swift\ 		& 00037050003	& XRT			& PC  		& 2009 Feb 25 05:15:40 -- 2009 Feb 25 08:34:56 	& 1.4 		& $0.066\pm0.007$		\\ \vspace{0.15cm}
\swift\ 		& 00037050004	& XRT			& PC  		& 2018 Feb 23 03:13:45 -- 2018 Feb 23 06:01:12 	& 1.7  		& $0.035\pm0.005$		\\  \vspace{0.15cm}
\swift\ 		& 00037050005	& XRT			& PC  		& 2018 Feb 28 03:49:52 -- 2018 Feb 28 05:46:53 	& 3.0			& $0.039\pm0.004$		\\ 
\xmm\   		& 0764344301   	& EPIC pn                 & FT                 & 2018 Jun 20 20:36:41 -- 2018 Jun 21 14:27:22	& 62.8               & $0.584\pm0.004$		\\ 
                         &                               & EPIC MOS\,1         & LW                 & 2018 Jun 20 19:56:41 -- 2018 Jun 21 14:27:03 	& 65.4               & $0.172\pm0.002$		\\ \vspace{0.15cm}
                         &                               & EPIC MOS\,2         & SW                 & 2018 Jun 20 19:59:25 -- 2018 Jun 21 14:29:28  	& 64.1               & $0.199\pm0.002$		\\                             
\nustar\      	& 80101101002        & FPMA                    & Science             & 2018 Jun 20 10:01:09 -- 2018 Jun 21 18:01:09	& 67.5          	& $0.065\pm0.001$	\\
                         &                               & FPMB                    & Science            &  2018 Jun 20 10:01:09 -- 2018 Jun 21 18:01:09    & 67.5          	& $0.060\pm0.001$	\\
\hline
\end{tabular}
}
\end{center}
{\bf Notes.}
PC stands for `photon counting', TE FAINT for `timed exposure' with faint telemetry format, FT for `fast timing', LW for `large window' and SW for `small window'. The average net count rates refer to the 0.3--10~keV energy band for \swift\ and \xmm\ data, to the 0.3--8~keV energy band for \cxo\ data, and to the 3--79~keV energy range for \nustar\ data.
\end{table*}

\subsection{XMM--Newton \emph{EPIC} and \emph{RGS}}

\src\ was observed with the European Photon Imaging Cameras (EPIC) and the Reflection Grating Spectrometer (RGS) arrays on board \xmm\ for about 68~ks on 2018 Jun 20--21 (see Table~\ref{tab:log}). The pn camera (Str\"{u}der et al. 2001) was set in fast timing (FT) mode (time resolution $\Delta t_{{\rm res}}\sim29.52$~$\mu$s) using the thin filter. The MOS1 and MOS2 cameras (Turner et al. 2001) were operated in large ($\Delta t_{{\rm res}}\sim0.9$~s) and small ($\Delta t_{{\rm res}}\sim0.3$~s) window modes, respectively, using the thin filter. The RGSs (den Herder et al. 2001) were operated in the standard spectroscopy mode ($\Delta t_{{\rm res}}\sim5.7$~s) throughout the observation. However, the RGS data were discarded from our analysis, as the low signal-to-noise ratio (S/N) precluded any meaningful study of the source X-ray emission properties.

We processed the raw EPIC observation data files within \texttt{SAS}, using the most up to date calibration files available. We extracted a light curve of single-pixel events (\texttt{pattern = 0}) for the whole field of view (FoV) and for each EPIC instrument, and filtered out the datasets corresponding to the initial $\sim3$~ks of the observation, as these were affected by a slightly enhanced background flaring activity. For the pn camera we extracted the source counts within a ten-pixel-wide strip orientated along the readout direction of the CCD and centred on the brightest column of CCD pixels (RAW-X = 37; one EPIC pn pixel corresponds to 4.1 arcsec). Background events were accumulated within a 3-pixel-wide region far from the position of the source ($3 \leq$ RAW-X $\leq 5$). For the two MOSs, source counts were collected within a circle with a radius of 30 arcsec, and the background from a circle of the same size located on the same CCD. The average source net count rate over the 0.3--10~keV energy band was $0.584 \pm 0.004$~counts~s$^{-1}$ for the pn, $0.172 \pm 0.002$~counts~s$^{-1}$ for the MOS1 and $0.199 \pm 0.002$~counts~s$^{-1}$ for the MOS2. For the timing analysis, we referred the photons' arrival times of all event lists to the Solar System barycentre reference frame using the DE-200 Solar System ephemeris (which we adopt in the following for all data sets). We extracted the light curve of \src\ over the 0.3--10~keV energy range by combining the background-subtracted and exposure-corrected time series from each camera during the periods when all three instruments acquired data simultaneously. For the spectral analysis, we retained only single and double pixel events optimally calibrated (\texttt{pattern} $\leq 4$, \texttt{flag} = 0). We produced the redistribution matrices and effective area files for each spectrum, and grouped the background-subtracted spectra following the optimal binning prescription of Kaastra \& Bleeker (2016). We restricted the spectral analysis of the pn data to energies above 0.7~keV, owing to known calibration issues of this instrument at low energy when operating in the FT mode.

\subsection{NuSTAR}

The \emph{Nuclear Spectroscopic Telescope Array} mission (\nustar; Harrison et al. 2013) observed \src\ on 2018 June 20--21 (obs. ID: 80101101002; PI: A. Papitto) for an elapsed time of 115.2~ks and an effective net on-source exposure time of about 67.5~ks. The observation began $\sim10$~h before and ended $\sim3.5$~h after the \xmm\ pointing, thus yielding complete simultaneity with the \xmm\ observation (see Table~\ref{tab:log}). We processed the data using the script \texttt{nupipeline} (v.~0.4.6) under the \nustar\ Data Analysis Software (\texttt{nustardas} v.~1.9.3\footnote{\texttt{nustardas} v.~1.9.3 is distributed along with \texttt{HEASOFT} v.~6.25.}), and adopted the most recent instrumental calibration files (v20180419). We also filtered the data excluding time intervals related to passages of the satellite through the South Atlantic Anomaly (SAA), and corrected for drifts of the spacecraft clock adopting the \nustar\ clock file v.~82. The combined image of the field around \src\ extracted using both focal plane modules (FPMA, FPMB) is shown in the inset of Fig.~\ref{fig:mosa}. For each FPM, we collected the source counts within a circular region of radius 50 arcsec, and background photons from a circle of radius 135 arcsec located on the same detector chip far away from the source. The average source net count rate over the 3--79~keV energy band was $0.065 \pm 0.001$~counts~s$^{-1}$ for the FPMA and $0.060 \pm 0.001$~counts~s$^{-1}$ for the FPMB. We referred the photons' times of arrival to the Solar System reference frame. Background-subtracted and exposure-corrected light curves, source and background spectra, response matrices and ancillary files were created separately for each FPM using \texttt{nuproducts} (v.~0.3.0). The light curves were combined to increase the S/N, and to probe the variability pattern of the source more in detail in the hard X-ray band. The spectra were grouped using the scheme of Kaastra \& Bleeker (2016), as for the \xmm\ data sets.

\subsection{ROSAT}

\src\ entered the FoV of the Position-Sensitive Proportional Counter (Pfeffermann et al. 1987) on board the {\em Roentgen} Satellite (\ros) for $\sim430$ s on 1991 January 18--21, during the satellite all-sky survey (sequence ID: RS932816N00). The source was detected at a count rate of ($27\pm9$)~counts~ks$^{-1}$ in the 0.1--2.4~keV energy band, computed applying the \texttt{sosta} tool of \texttt{Ximage} (v.~4.5.1) on the processed image.

\subsection{Swift \emph{XRT}}

\src\ was observed three times with the X-ray Telescope (XRT; Burrows et al. 2005) on board \swift\ (Gehrels et al. 2004) in 2008 and 2009. We requested two additional pointings more recently, on 2018 February 23 and 28, to assess the source flux after a decade-long gap of X-ray observations of the field, and to unveil possible long-term spectral and flux variations of the soft X-ray emission. In all cases, the XRT was configured in the photon counting (PC) mode, yielding a readout time of $\sim2.5$~s (see Table~\ref{tab:log}).

We processed all data with standard screening criteria and generated exposure maps with the task \texttt{xrtpipeline} (v.~0.13.4) from the \texttt{FTOOLS} package. We extracted the source and background event lists and spectra using a circular region with a radius of 20 pixels (corresponding to 90\% of the XRT encircled energy fraction at 1.5 keV) and an annulus with inner and outer radius of 40 and 80 pixels, respectively (one XRT pixel corresponds to about 2.36 arcsec). For all the observations we extracted exposure-corrected and background-subtracted light curves using \texttt{xrtlccorr} and \texttt{lcmath} (the latter accounting also for different areas of the source and background extraction regions). We created the observation-specific ancillary response files, assigned the appropriate redistribution matrix available in the \texttt{HEASARC} calibration data base, and filtered out photons with energy below 0.3~keV. 
Because some of the spectra were characterised by too few counts for the $\chi^2$-fitting, we decided to group each dataset so as to contain at least 5 counts in each spectral bin, and used the Cash statistics (Cash 1979).

\subsection{Chandra}

The \emph{Chandra} X-ray Observatory observed the field of \src\ with the imaging array of the Advanced CCD Imaging Spectrometer (ACIS-I; Garmire et al. 2003) for $\simeq5$~ks on 2008 September 12, as part of a programme aimed at localising and characterising the soft X-ray counterparts of unclassified \integ\ sources (Tomsick et al. 2009). The ACIS-I was set in full-imaging timed-exposure mode (frame time of 3.24~s) with faint telemetry format. 

We re-analysed the data following the standard analysis threads for a point-like source with the \cxo\ Interactive Analysis of Observations software (\texttt{CIAO}, v.~4.10; Fruscione et al. 2006) and the calibration files stored in the \cxo\ CALDB (v.~4.8.1). First, we reprocessed the data using the \texttt{chandra$_{-}$repro} script. The exposure-corrected image of the sky region around \src\ is shown in Fig.~\ref{fig:fov}. \src\ was located far off-axis, about 5.8 arcmin away from the position of the aim point. Because the shape and size of the ACIS point spread function are known to change significantly as a function of the off-axis angle, we decided to apply the \texttt{psfsize$_-$srcs} tool to estimate the radius of the 90\% encircled counts fraction in the 0.5--7.0~keV energy band at a similar offset angle. Source photons were hence collected within a circular region with a radius of 5 arcsec. The background was extracted from an annulus centred on the source location, with inner and outer radius of 6 and 12 arcsec, respectively. We visually inspected the time series of the background emission, and did not find any flaring episode during the whole duration of the observation, regardless of the adopted time bin. We hence retained all datasets in the following analysis. We then restricted the analysis to the 0.3--8 keV energy interval, where the calibration of the spectral responses is best known. The average source net count rate in this energy band was $0.182 \pm 0.006$~counts~s$^{-1}$. The times of arrival of the source photons were referred to the Solar System reference frame. The source and background spectra, the associated redistribution matrix and ancillary response file were generated by means of the \texttt{specextract} script. The background-subtracted spectrum was rebinned as explained by Kaastra \& Bleeker (2016).

\subsection{XMM-Newton slew observations}

The field of \src\ was covered by \xmm\ slew observations on 2012 February 4, 2014 February 11 and 2017 January 5, when the spacecraft was moving between scheduled pointing positions during revolutions 2226, 2596 and 3127, respectively (see Table~\ref{tab:slew} for a journal of the observations). The slew observations are based on data from the pn camera alone, with the medium optical blocking filter in place. The in-orbit slew speed of 90\deg~h$^{-1}$ translates into exposure times in the 1--11~s range, yielding a sensitivity limit on the observed flux of $\sim1.2 \times 10^{-12}$ \flux\ over the 0.2--12~keV energy interval (see Saxton et al. 2008). We ran the \texttt{eslewchain} and \texttt{eslewsearch} tasks of \texttt{SAS} to process the data and compile a list of detected sources, respectively\footnote{\texttt{https://www.cosmos.esa.int/web/xmm-newton/\\sas-thread-epic-slew-processing}}. \src\ was detected at a count rate of $0.7\pm0.3$~counts~s$^{-1}$, $0.6\pm0.3$~counts~s$^{-1}$ and $0.9\pm0.4$~counts~s$^{-1}$ over the 0.2--12~keV energy band in 2012, 2014 and 2017, respectively (see Table~\ref{tab:slew}). Hence, the count rate was consistent with being the same within the uncertainties along the three epochs.

\subsection{INTEGRAL}
\label{sec:integral}

The field around \src\ was observed with the International Gamma-Ray Astrophysics Laboratory (\integ; Winkler et al. 2003) from 2003 January 29 to 2018 February 28, covering the satellite revolutions 36--1916. We processed and analysed all the publicly available data using the Off-line Scientific Analysis software (\texttt{OSA} v.~10.2) distributed  by the \integ\ Science Data Centre (Courvoisier et al. 2003). 

The \integ\ observations are divided into "science windows" (SCWs), i.e. pointings with typical durations of $\sim$2-3~ks. For the analysis we retained all SCWs for which the source was located within an off-axis angle from the centre of the instrument FoV of 12\deg\ for the ISGRI detector (Lebrun et al. 2003) on the IBIS instrument (Ubertini et al. 2003) and 3.5\deg\ for the JEM-X (Lund et al. 2003). These choices allowed us to minimise the instruments calibration uncertainties\footnote{\texttt{http://www.isdc.unige.ch/integral/analysis}}, and led to a selection of a total of 4078 SCWs. 

\begin{figure*}
\begin{center}
\includegraphics[width=1.1\textwidth]{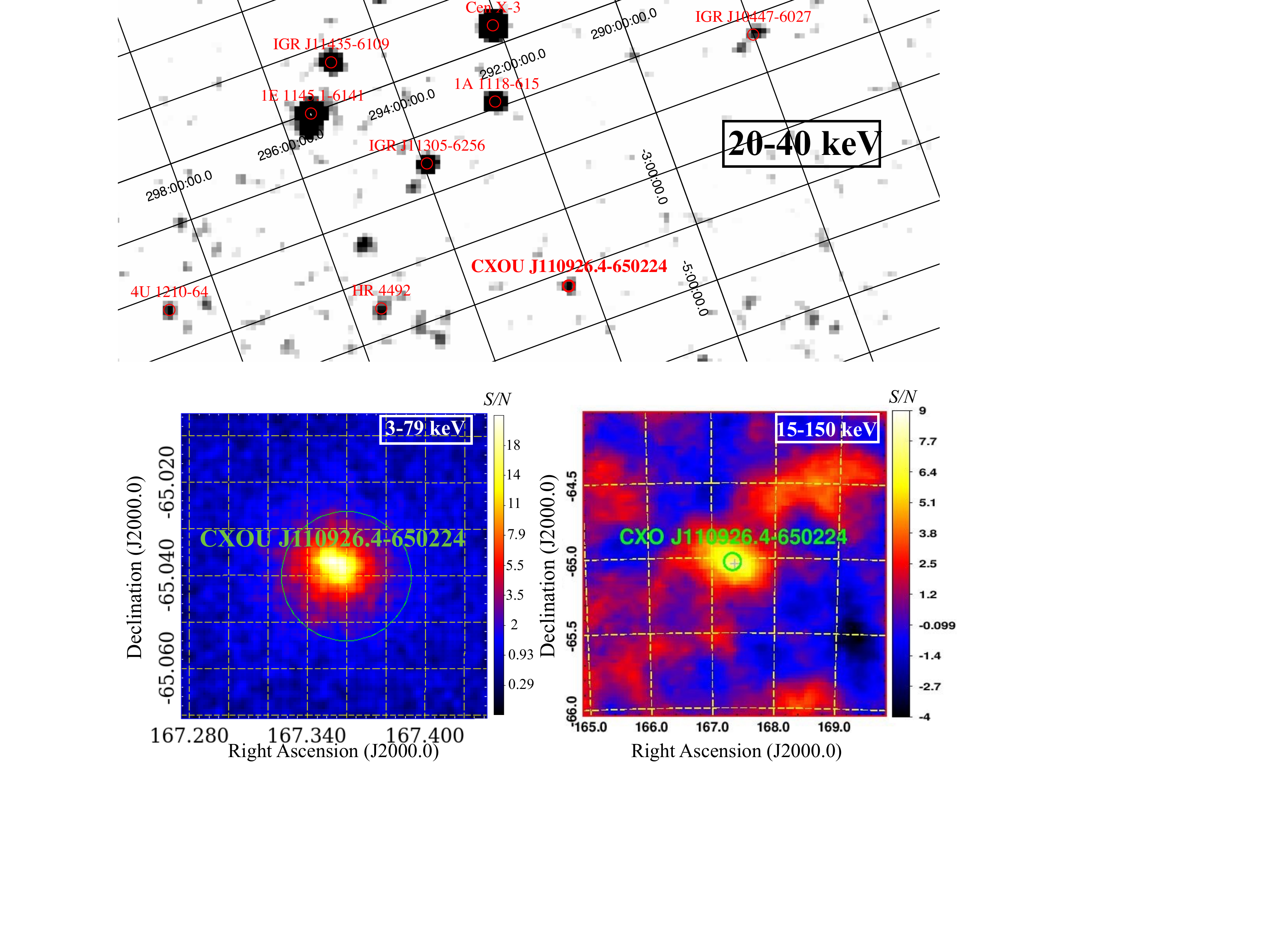}
\vspace{-3.2cm}
\caption{Images of the sky region around \src\ in different broad X-ray energy ranges obtained using different instruments. {\it Top panel}: \integ\ IBIS/ISGRI 20--40~keV mosaicked image. Galactic coordinates are reported. The grey scale is proportional to the X-ray flux. All sources detected in the field are labelled (\src\ is labelled in bold). {\it Lower panels, left}: \nustar\ FPMA$+$FPMB 3--79~keV image with sub-pixel binning. The 50-arcsec circle adopted to extract the source photons is overlaid. The colour code indicates the S/N at the different positions in the image. {\it Lower panels, right}: \swift\ BAT 15--150~keV image. The colour code indicates the S/N at the different positions in the image (negative values are in logarithmic scale). Different scales have been adopted for the images in the lower panels (more enlarged from left to right).
 }
\label{fig:field_bat}
\label{fig:fermi_map}
\label{fig:mosa}
\end{center}
\end{figure*}

We extracted the IBIS/ISGRI and JEM-X mosaic images by combining the images acquired in the single SCWs. \src\ was detected in the IBIS/ISGRI 20--40~keV mosaic (effective exposure of 6962~ks) at a significance of $6.9\sigma$ and average count rate of $0.07\pm0.01$~counts~s$^{-1}$ (this corresponds to $0.5\pm0.1$~mCrab, i.e. roughly 4$\times$10$^{-12}$~\flux). The source was instead not detected by JEM-X, and we estimated a $3\sigma$ upper limit on its 3--20~keV flux of 0.6~mCrab (roughly 1.4$\times$10$^{-11}$~\flux) from the mosaic. The upper panel of Fig.~\ref{fig:mosa} shows a zoom of the IBIS/ISGRI mosaic of the region around the source.

Given the relatively low detection significance in the IBIS/ISGRI mosaic, we extracted for this instrument a single spectrum integrating over the entire exposure time available. A power-law fit to the 20--50~keV spectrum yielded a photon index of $\Gamma=3.3\pm1.9$ and a flux $\sim6\times10^{-12}$~\flux\ (reduced chi-squared $\chi^2_{{\rm red}} = 0.3$ for 3 degrees of freedom, d.o.f. hereafter). In Sect.~\ref{sec:xraylongterm} we will fit again the spectrum by fixing the photon index to the better constrained value measured by the joint spectral fits of the \xmm\ and \nustar\ datasets. 

We also inspected all IBIS/ISGRI SCWs to search for possible few ks-long flares from the source. We extracted the light curves with a time resolution of 1 SCW over the 20--40~keV, 40--80~keV and 20--100~keV energy bands, but found no significant detections.

\begin{table}
\begin{center}
\caption{Log of the \xmm\ EPIC pn slew observations of \src.} 
\label{tab:slew}
\resizebox{1\columnwidth}{!}{
\begin{tabular}{clccc} 
\hline \hline 
Slew ID				&  Date  			& Exposure  	& Net count rate 	& Detection 		\\
         				& 				& (s)   		& (counts~s$^{-1}$)	& likelihood		\\
\hline 			
9222600004			& 2012 Feb 4   		& 10  		& $0.7\pm0.3$		& $\sim25$	\\ 
9259600004			& 2014 Feb 11 		& 7			& $0.6\pm0.3$		& $\sim17$	\\ 
9312700008			& 2017 Jan 5   		& 6  			& $0.9\pm0.4$		& $\sim19$	\\ 
\hline
\end{tabular}
}
\end{center}
{\bf Notes.} The pn was configured in the full frame mode in all cases. The average net count rate refers to the 0.2--12~keV energy band.
\end{table}

\subsection{Swift \emph{BAT}}

The field has been continuously monitored in the hard X-ray range also by the BAT (Barthelmy et al. 2005) on board \swift, since the satellite launch. The hard X-ray counterpart of \src, also listed in the fourth Palermo BAT catalogue with the name of 4PBC\,J1109.3$-$6501, was detected at an average count rate of $(1.4\pm0.2)\times 10^{-5}$~counts~s$^{-1}$ over the 15--150~keV energy interval between 2004 December and 2017 October, at a S/N of 7.3 (see the inset in Fig.~\ref{fig:field_bat}). We downloaded the eight-channel average spectrum, the response matrix file and the light curve binned at 360~d from the archive  publicly available at the Palermo BAT survey website. 

The spectrum is well described by a power law with photon index of $\Gamma=1.6\pm0.3$ ($\chi^2_{{\rm red}} = 1.44$ for 6 d.o.f.), yielding a flux of $\sim7.8\times10^{-12}$~\flux\ over the 15--150~keV energy interval. The value for the photon index is compatible within the uncertainties with that derived from the spectral fit of the \integ\ data over a narrower energy interval (see Sect.~\ref{sec:integral}). In Sect.~\ref{sec:xraylongterm} we will fit again the spectrum by fixing the photon index to the value measured with the joint \xmm\ and \nustar\ spectral fits.

\section{Radio observations}
\label{sec:radio}

We observed the field around \src\ with the Australia Telescope Compact Array (ATCA; Frater et al. 1992) on 2018 June 20 between 10:16:12 UTC and 15:42:36 UTC simultaneously at central frequencies of 5.5 and 9.0\,GHz. The data were recorded utilising the Compact Array broad-band Backend (CABB; Wilson et al. 2011), comprised of 2048 1-MHz channels. The total on-source integration time was $\sim 4.8$~h, entirely overlapping with the \nustar\ observation.

Data were processed and imaged following standard procedures within the Common Astronomy Software Applications package (\texttt{CASA} version 4.7.2; McMullin et al. 2007). We used 1934-658 as the primary bandpass and flux calibrator, and 1059-63 ($\sim$1.8\deg\ away) as the phase calibrator. Images were produced using a Briggs robustness parameter of 2 (natural weighting) to maximise sensitivity. With the array in the 1.5A array configuration, this choice provided angular resolutions of $6.0$ arcsec $\times$ 2.5 arcsec at 5.5\,GHz and $3.6$ arcsec $\times$ 1.4 arcsec at 9.0\,GHz, with a position angle of 68\deg\ (measured from north towards east).

\section{Results}
\label{results}

\subsection{The optical emission}
\label{sec:optprop}

The optical counterpart was detected in all the SALTICAM frames (see Table~\ref{tab:phot}). The emission was highly and rapidly variable, increasing by $\sim0.3$~mag and $\sim0.4$~mag in the $g'$ and $r'$ filters, respectively, on timescales of hundreds of seconds. \src\ was also only detected in two consecutive \xmm\ OM images  (out of seven exposures), at a magnitude $B=20.47\pm0.09$ and $B=20.45\pm0.09$ in the Vega system. This shows that the optical variability extends to longer timescales, of the order of a few ks. Interestingly, these two detections correspond to time intervals when flaring activity was registered in the X-ray band during the last part of the observation (see Sect.~\ref{sec:xraytiming}), possibly suggesting a correlated variability pattern. However, the low S/N in the two OM images precluded any meaningful characterization of the optical variability and of a possible correlation between the optical and X-ray emissions.

\begin{figure*}
\centering
\includegraphics[width=0.99\textwidth]{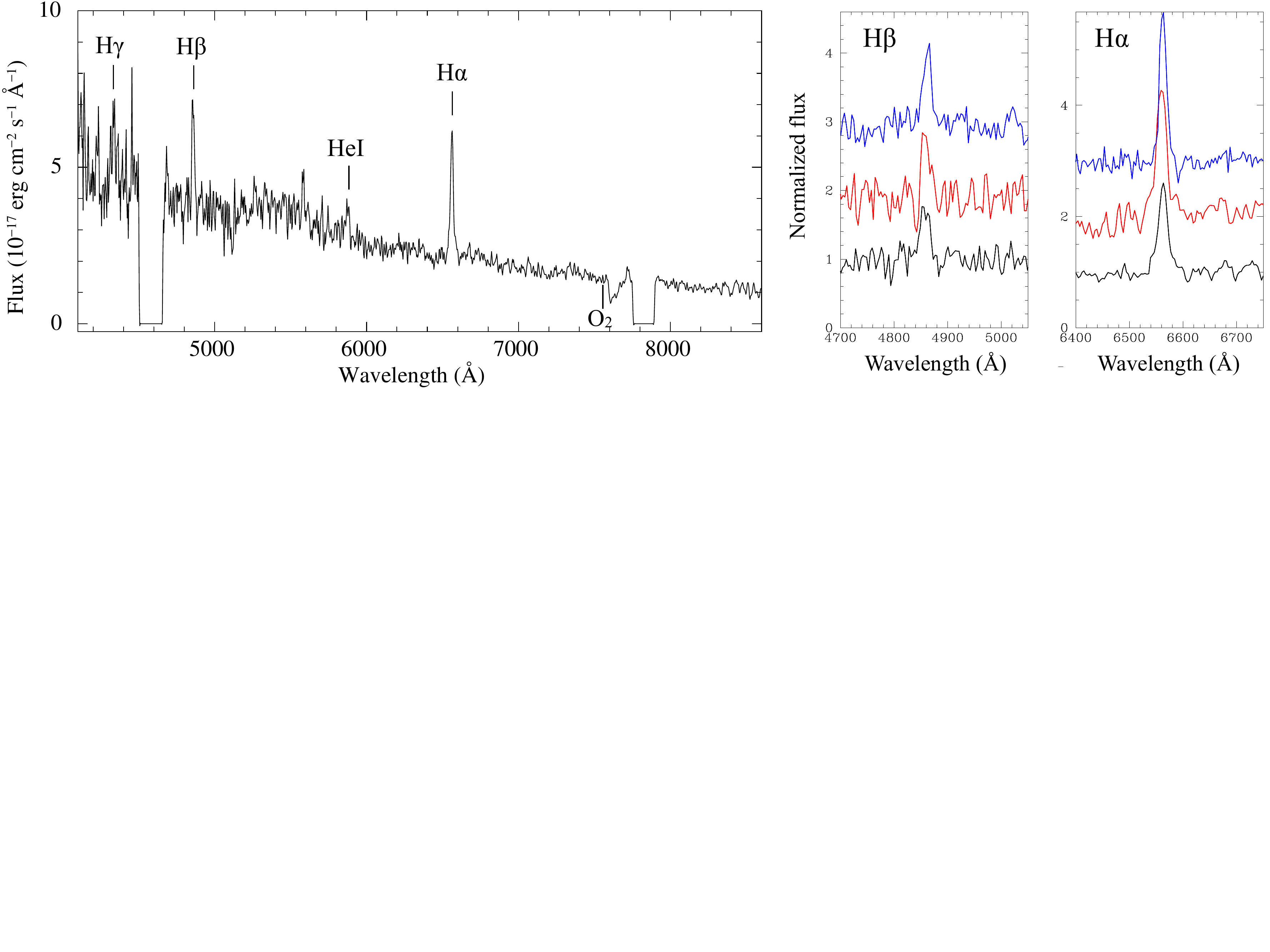}
\vspace{-7.9cm}
\caption{{\it Left-hand panel}: low-resolution optical spectrum of \src\ obtained from the average of the two spectra acquired on 2018 March 4 using SALT RSS (total exposure of 2000~s), and restricted to the 4100--8600~\AA\ wavelength range. The gaps over the 4530--4630~\AA\ and 7790--7860~\AA\ wavelength ranges are due to the gaps between the CCD chips in the RSS. The most prominent emission features as well as the telluric (atmospheric) O$_2$ absorption feature are labelled. {\it Right-hand panel}: zoom on the wavelength ranges around the H$\beta$ and H$\alpha$ emission lines for the three spectra (in black, red and blue following the chronological order of spectra acquisition). Spectra were normalised to the continuum, and shifted along the vertical axis for better visualization.}
\label{fig:opt_spec}
\end{figure*}

All three spectra showed prominent, broad Balmer emission lines with variable profiles in time (see Fig.~\ref{fig:opt_spec} and Table~\ref{tab_EW}). On the other hand, the low-resolution spectra did not allow us to identify possible absorption features of the donor star.

\begin{table}
\begin{center}
\caption{Journal of the photometric observations of \src\ performed with SALTICAM on 2018 March 4.} 
\small
\label{tab:phot}
\begin{tabular}{ccc} \hline \hline 
SDSS filter	& Start time  	& Magnitude \\ 
			& (UTC)		& \\
\hline		
$r'$			& 00:52:29.33	& $19.90\pm0.02$  \\ 
$g'$			& 00:54:49.53 	& $20.74\pm0.03$ \\	
$g'$			& 00:58:56.95	& $20.44\pm0.01$ \\
$i'$			& 01:01:34.63	& $19.08\pm0.06$  \\	
$r'$ 			& 01:08:19.93	& $19.52\pm0.01$ \\ 
$g'$ 			& 01:12:47.80	& $20.53\pm0.01$ \\
				
\hline
\end{tabular}
\end{center}
{\bf Notes.} Each exposure lasted 100~s. Magnitudes are not corrected for intervening absorption along the line of sight towards \src.
\end{table}

\begin{table}
\centering
\small
\caption{Properties of the Balmer emission lines of \src\ measured for the three spectra acquired with SALT RSS (see the text for details). Numbers in parentheses denote the uncertainty on the last quoted digit.}
\resizebox{.95\columnwidth}{!}{
\begin{tabular}{cccc c ccc} \hline  \hline 
Line           		&  \multicolumn{3}{c}{Equivalent width (\AA)}    &  \multicolumn{3}{c}{Full-width at half-maximum (\AA)}                \\
\cmidrule(lr){2-4} \cmidrule(lr){5-7}
               			&  Spec1 		& Spec2 		& Spec3          	&  Spec1 		& Spec2 		& Spec3     \\ \hline \vspace{0.1cm}
H$\gamma$     		& $8(1)$ 		& $5.0(1)$ 	& $21(1)$    	& $30(5)$	 	& $7(2)$ 		& $56(3)$  	   \\ \vspace{0.1cm}
H$\beta$      		& $15(1)$ 		& $22(2)$ 		& $16(3)$   	& $24(1)$ 		& $16(3)$	 	& $15(1)$  	   \\ \vspace{0.1cm}
H$\alpha$      		& $32(1)$ 		& $45(5)$	 	& $48(6)$    	& $18.2(1)$ 	& $20(2)$	 	& $14.8(2)$  	   \\
\hline
\end{tabular}  
}
\label{tab_EW}
\end{table}

\begin{figure*}
\includegraphics[width=1.31\textwidth]{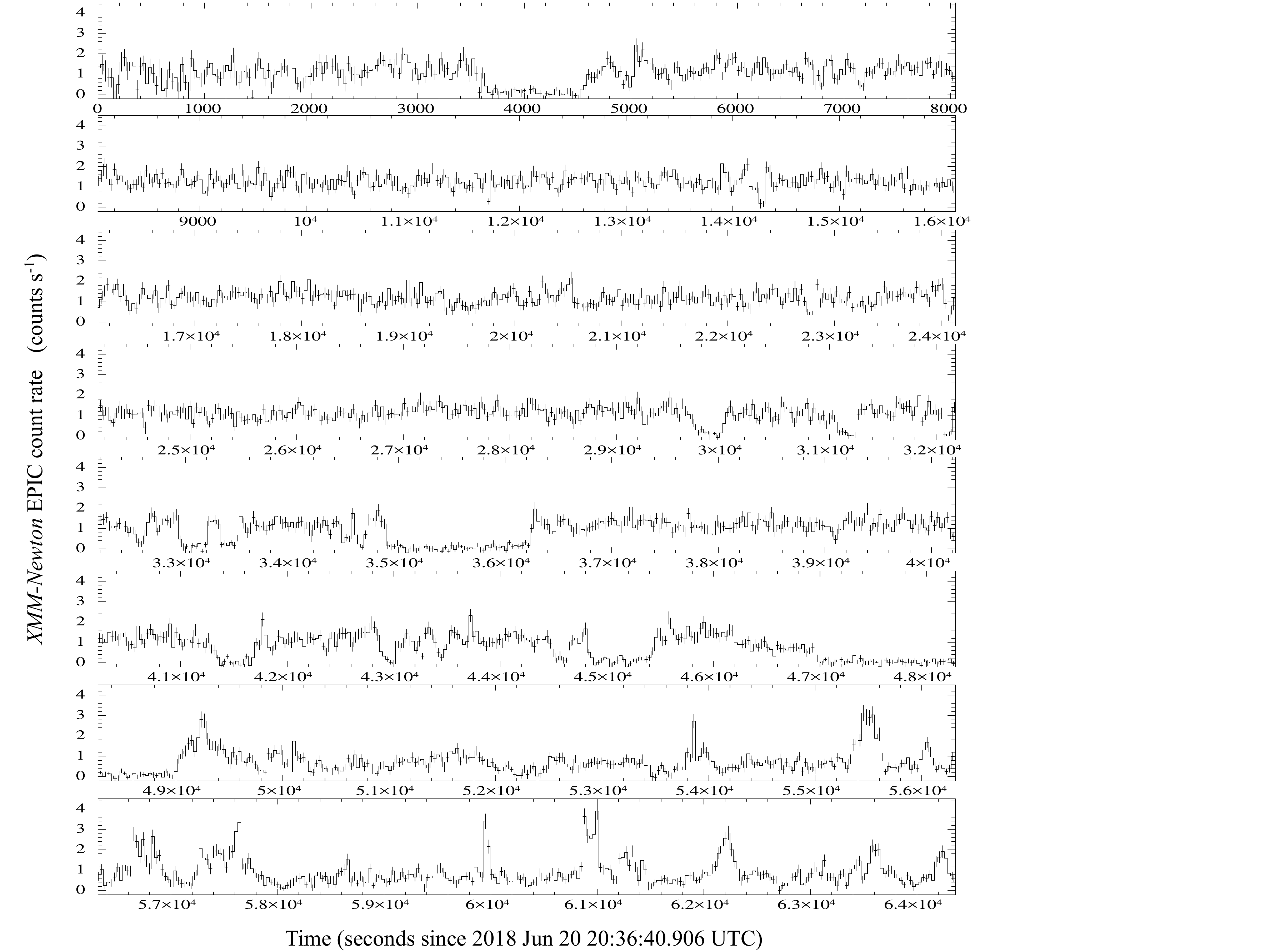}
\vspace{0.05cm}
\caption{0.3--10 keV background-subtracted and exposure-corrected light curve of \src\ obtained with \xmm\ EPIC with a time bin of 30~s.}
\label{fig:xmm_lc}
\end{figure*}

\begin{figure*}
\begin{center}
\includegraphics[width=1\textwidth]{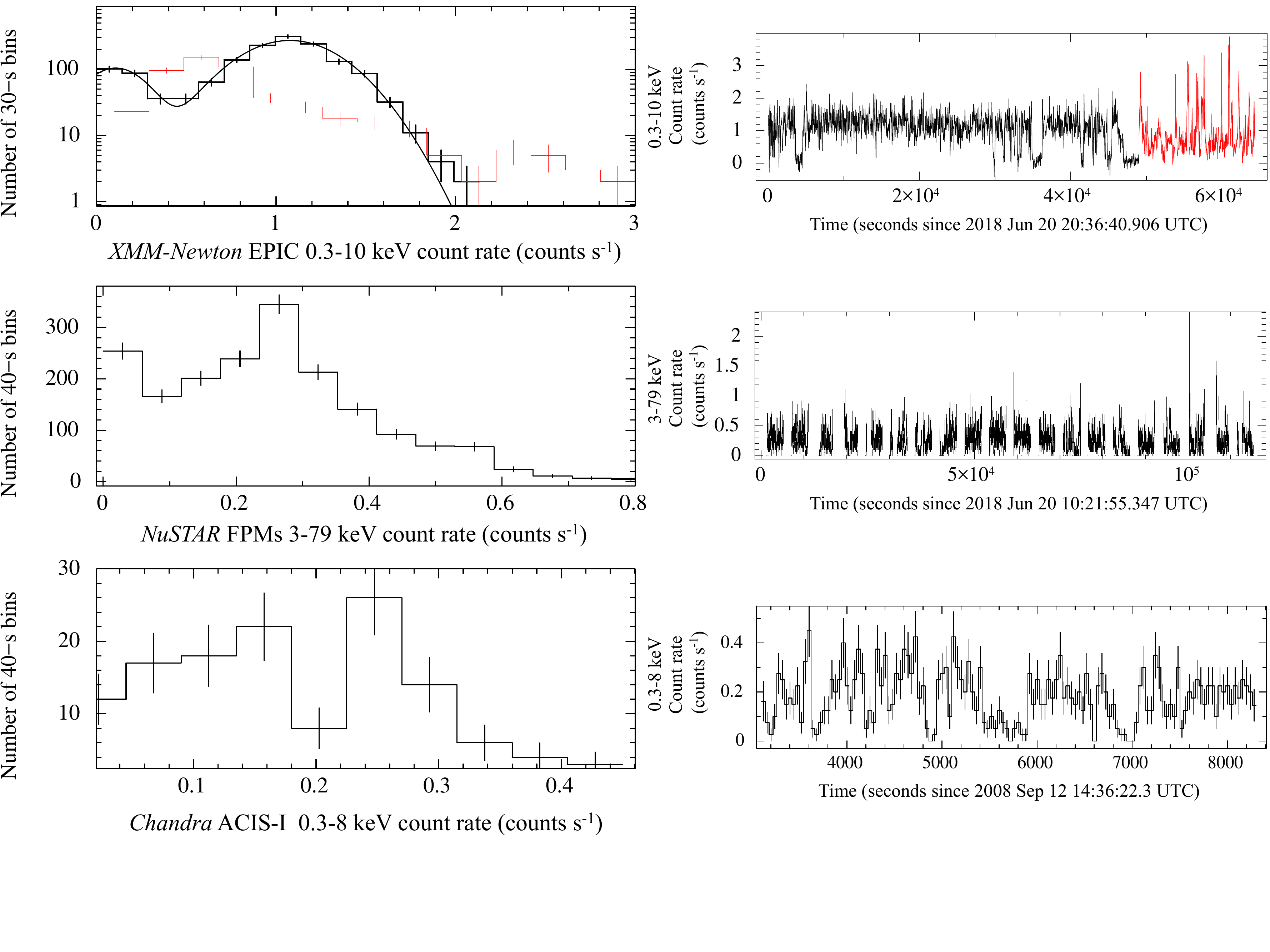}
\vspace{-1.8cm}
\caption{Distribution of the count rates ({\it left-hand panels}) obtained from the binned background-subtracted and exposure-corrected light curves of \src\ extracted using different X-ray instruments ({\it right-hand panels}). The vertical error bars in the distributions are evaluated as the square root of the number of time bins measured in the corresponding count rate interval (i.e. as the standard deviation of a Poisson distribution). {\it Top panel}: from the \xmm\ EPIC 30-s binned light curve, separately for the first 49~ks (in black) and the last 16~ks (in red) of the observation. The black solid line represents the best-fitting model for the former distribution, i.e. the superposition of two log-normal distributions (see the text for details). A logarithmic scale was adopted for the vertical axis, for better visualization. {\it Middle panel}: from the \nustar\ FPMA$+$FPMB 40-s binned light curve. The $\sim3$-ks long gaps in the light curve are due to Earth occultation. {\it Bottom panel}: from the \cxo\ ACIS-I 40-s binned light curve.}
\label{fig:distrib_cxo}
\end{center}
\end{figure*}

\subsection{The X-ray emission}
\label{sec:xrayresults}

\subsubsection{Timing properties}
\label{sec:xraytiming}

The light curve extracted from \xmm\ EPIC data is reported in Fig.~\ref{fig:xmm_lc}. It shows unpredictable, substantial, rapid changes in the net count rate on timescales as short as 30~s. A bimodal trend of the net count rates was particularly evident over the first $\sim49$~ks of the observation: during a few time instances, the net count rate dropped sharply to zero for elapsed times ranging from a minimum of a few tens of seconds up to about half a hour (e.g. during the time intervals 3.8--4.5~ks, 35.0--36.2~ks and 47--49~ks in Fig.~\ref{fig:xmm_lc}). These flat-bottomed dips showed asymmetric profiles with a variety of timescales for the changes of the net count rates at the beginning and at the end of the dips, between tens of seconds and about ten minutes. Following a prolonged dipping episode lasting about half a hour, the source underwent a rather powerful flare (during the time interval 49.0--49.8~ks in Fig.~\ref{fig:xmm_lc}) with a peak count rate of $\sim2.8$~counts~s$^{-1}$. It then entered a period of erratic (i.e., no more bimodal) variability characterised by a lower average count rate and sporadic flaring episodes superposed. These episodes exhibited different profiles (e.g. saw-toothed, triangular, flat-top, mono- and double-peaked) and variable duration in the range $\sim1-10$~min, and reached slightly different peak intensities. 

This change in the emission pattern can be visualised more clearly in the top-left panel of Fig.~\ref{fig:distrib_cxo}, where we report the distribution of the count rates separately for the first and second part of the observation. The morphologies of the two distributions are clearly different. The former (black colour in the figure) displays a double-peaked profile, which is adequately described by the superposition of two log-normal distributions with centroids at $0.11_{-0.06}^{+0.03}$~counts~s$^{-1}$ and at $1.077\pm0.008$~counts~s$^{-1}$ for the secondary peak and the more prominent peak, respectively ($\chi^2_{{\rm red}} = 1.2$ for 9 d.o.f.). Based on the $F$-test, the inclusion of a second log-normal distribution component yielded a null-hypothesis probability of $\simeq 3 \times 10^{-5}$. The latter (red colour in the figure) is instead characterised by a single peak that is located in between the two peaks recorded during the first part of the observation, at a value of $\simeq0.6$~counts~s$^{-1}$. The distribution decays smoothly with increasing count rate, and an additional, separate tail appears at $\gtrsim 2.2$~counts~s$^{-1}$. It is related to the above-mentioned flaring activity (no flares were instead registered over the first 49~ks of the observation).  

A very similar variability pattern on the same timescales was observed at higher energy in the nearly simultaneous \nustar\ datasets, as well as in archival \cxo\ datasets. In particular, the \nustar\ data showed that \src\ maintained the state of erratic variability (and underwent a few more flares) at least until the end of the observation, hence for at least $\sim12.7$~ks after the end of the \xmm\ pointing. We show the distribution of count rates derived from the \nustar\ and \cxo\ data and the corresponding light curves in the middle and bottom panels of Fig.~\ref{fig:distrib_cxo}, respectively. The histograms display a bimodal shape, although the relatively scarce photon counting statistics precludes a detailed sampling of the count rate distribution as for the \xmm\ EPIC data.

We did not find any evidence for the presence of a recurrence time between consecutive dips or flares. Moreover, we did not observe any (anti-)correlation between the dip duration and the time elapsed before the occurrence of the following dip, or between the peak intensity reached during flares and their fluence, rise and/or decay time or overall duration. 

\begin{figure}
\begin{center}
\includegraphics[width=0.485\textwidth]{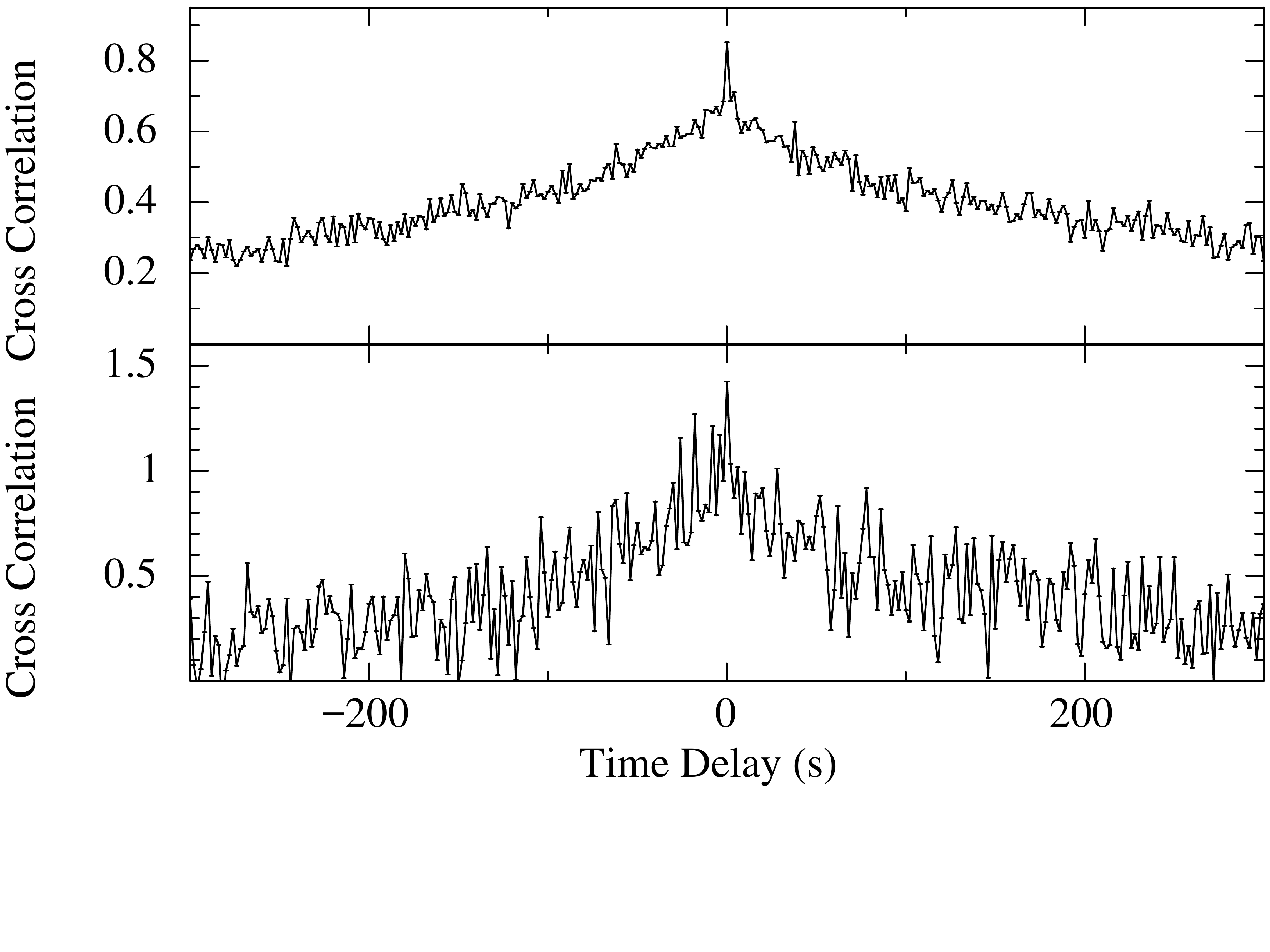}
\vspace{-1.3cm}
\caption{$z$-transformed discrete correlation functions (\emph{z}DCFs) between the light curves of \src\ extracted in the soft and hard X-ray energy bands from \xmm\ EPIC pn ({\it top panel}) and \nustar\ FPMA+FPMB ({\it bottom panel}) data using a time step of 2~s (see Sect.~\ref{sec:xraytiming} for the definitions adopted for the energy bands). The \emph{z}DCFs have been normalised so as to account for Poisson noise. Positive time delays correspond to higher energy photons lagging lower energy photons. For plotting purpose, an extended range of time delays is shown, and uncertainties on the values have been omitted.}
\label{fig:crosscor}
\end{center}
\end{figure}

To investigate correlations and lags between the soft and hard X-ray emissions, we extracted the light curves separately over the 0.3--2~keV and 2--10~keV energy ranges for the \xmm\ EPIC data sets, and over the 3--7.5~keV and 7.5--79~keV energy intervals for the \nustar\ data. These energy ranges were selected so as to achieve a comparable number of source net counts in both bands. We computed the $z$-transformed discrete correlation function ($z$DCF) between the light curves adopting the algorithm developed by Alexander (1997), which is based upon the DCF method of Edelson \& Krolik (1988). It is particularly effective for unevenly sampled datasets, as those acquired by \nustar. The $z$DCFs are reported in Fig.~\ref{fig:crosscor}, and display a symmetric shape with a prominent peak at zero lag. We calculated the uncertainty on the peak width using the maximum likelihood method by Alexander (2013), and found the peak location to be consistent with absence of time lag (at 1$\sigma$ c.l.)\footnote{We used the codes available at \texttt{http://www.weizmann.ac.il/\\particle/tal/research-activities/software}}.

Hereafter we define three different flux modes taking the EPIC 0.3--10~keV light curve as a baseline. We adopted the following selections for the count rate ranges: $>2.2$~counts~s$^{-1}$ over the whole duration of the observation for the flare mode (leading to a total of eight distinct flaring episodes); [0.8--1.8]~counts~s$^{-1}$ and $<0.4$~counts~s$^{-1}$ during the first 49~ks of the observation for the high and low modes, respectively (see again Fig.~\ref{fig:distrib_cxo}). We checked that slightly different choices for the count rate thresholds yielded very similar results in the following analyses. Based on these choices, we estimated that \src\ spent fractional times of $\sim2$, $\sim63$ and $\sim12$\% in the flare, high and low modes, respectively, during the \xmm\ observation. The source spent also a non-negligible amount of time in a `transition' region, during which it switched back and forth across the different modes and hence attained values for the count rate outside the boundaries defined above. The lack of time delays between the soft and hard X-ray emissions (see above) led us to apply the good time interval files associated to the three modes in the EPIC data directly to the strictly simultaneous \nustar\ event files, and single out the flux modes over a broader energy interval. The resulting exposure times and count rates for the X-ray modes and the different X-ray instruments are listed in Table~\ref{tab:statelog}.

We computed a power density spectrum (PDS) over the whole EPIC pn observation length to search for periodicities having a probability smaller than $p=2.7\times10^{-3}$ (corresponding to a 3$\sigma$ c.l.) of being compatible with counting noise. No signal was detected over the 0.3--10~keV energy range, and we set an upper limit of $A<4.5$\% (3$\sigma$ c.l.) on the root mean squared (rms) amplitude for any coherent signal with frequency lower than 1.5~kHz. We notice, however, that tMSPs in the sub-luminous accretion disk state showed a coherent X-ray signal only during the high mode (Archibald et al. 2015; Papitto et al. 2015). Moreover, we point out that a coherent X-ray signal searched over the whole 17~h-long observation could be easily washed out if the putative pulsar belongs to a tight binary system with an orbital period of a few hours (as for the case of tMSPs), owing to Doppler frequency shifts induced by the motion of the compact object along the orbit. Hence, we carried out a more honed search for periodicities selecting the time intervals corresponding to the high mode, and accounting for the effects of the unknown binary orbital motion as follows. First, we performed a search over time intervals of length $\Delta t = 495$~s, which represents the optimal choice for a circular orbit, a sinusoidal signal with a period of 2.5~ms, an orbital period of 12~h, a mass for the donor star of 0.3~$M_{\sun}$ and a system inclination of $i\sim45$\deg\ (see Eq.~[21] by Johnston \& Kulkarni 1991). We derived a very loose upper limit of $A<67$\% on the rms pulse amplitude. We obtained a more constraining limit of $A<41$\% considering time intervals of length $2\Delta t$. Shorter time intervals should be considered to search effectively for a signal from a faster-spinning pulsar in a tighter orbit ($\Delta t \propto P^{1/2}_{\rm s}P^{2/3}_{\rm orb}$, where $P_{\rm s}$ is the spin period of the pulsar and $P_{{\rm orb}}$ is the binary orbital period). However, the count rate of \src\ ($\simeq 0.5$~counts~s$^{-1}$) is so low that a search over time intervals $\Delta t < 495$~s would be meaningless, as the amplitude sensitivity would be $\geq 100\%$. We then applied a quadratic coherence recovery technique (Wood et al. 1991; Vaughan et al. 1994). This is an acceleration search method that relies on the assumption that the sinusoidal Doppler modulation of the photons' times of arrival caused by the motion along a circular orbit can be approximated well by a parabola over a restricted time interval. We considered time intervals of length $2\Delta t$, i.e. the most effective for a spin period of 2.5~ms, an orbital period of 2.4~h, a mass for the donor star of 0.3~$M_{\sun}$ and a system inclination of $i\sim45$\deg\ (see Eq.~[22] by Johnston \& Kulkarni 1991). We corrected the times of arrival in each time interval using the relation $t'=\alpha t^2$, where 
\begin{equation}
\alpha=-\frac{1}{2}~\frac{a~{\rm \sin}(i)}{c}\Bigg(\frac{2\pi}{P_{\rm s}}\Bigg)\Bigg(\frac{2\pi}{P_{{\rm orb}}}\Bigg)^2~{\rm sin}(\phi_0),
\end{equation}
$a~{\rm sin} (i)/c$ is the projected semi-major axis of the NS orbit and $\phi_0$ is the orbital phase at the beginning of the time series. We varied the coefficient of the quadratic term in steps equal to $\delta\alpha=2.4\times10^{-10}$~s$^{-1}$ over the range $-1.4\times10^{-7}~{\rm s}^{-1} < \alpha < 1.4\times10^{-7}~{\rm s}^{-1}$, i.e. we performed $N_{\alpha} = 2\alpha_{max}/\delta\alpha = 1167$ quadratic time transformations on each time interval (the step and the range were computed using Eq.~[4] by Vaughan et al. 1994, assuming the values for the spin and orbital periods quoted above, as well as masses of 1.4~$M_{\sun}$ and 0.3~$M_{\sun}$ for the NS and the donor star, respectively). We extracted a PDS from all the corrected time series, and evaluated the detection level corresponding to a 3$\sigma$ c.l. taking into account the number of quadratic time transformations in each time interval, the number of time intervals considered, and the number of independent frequencies examined in the PDS. We did not detect any periodic signal above this threshold, and set an upper limit of $A<42$\% (3$\sigma$ c.l.) on the rms pulse amplitude.  

\begin{figure}
\begin{center}
\includegraphics[width=0.48\textwidth]{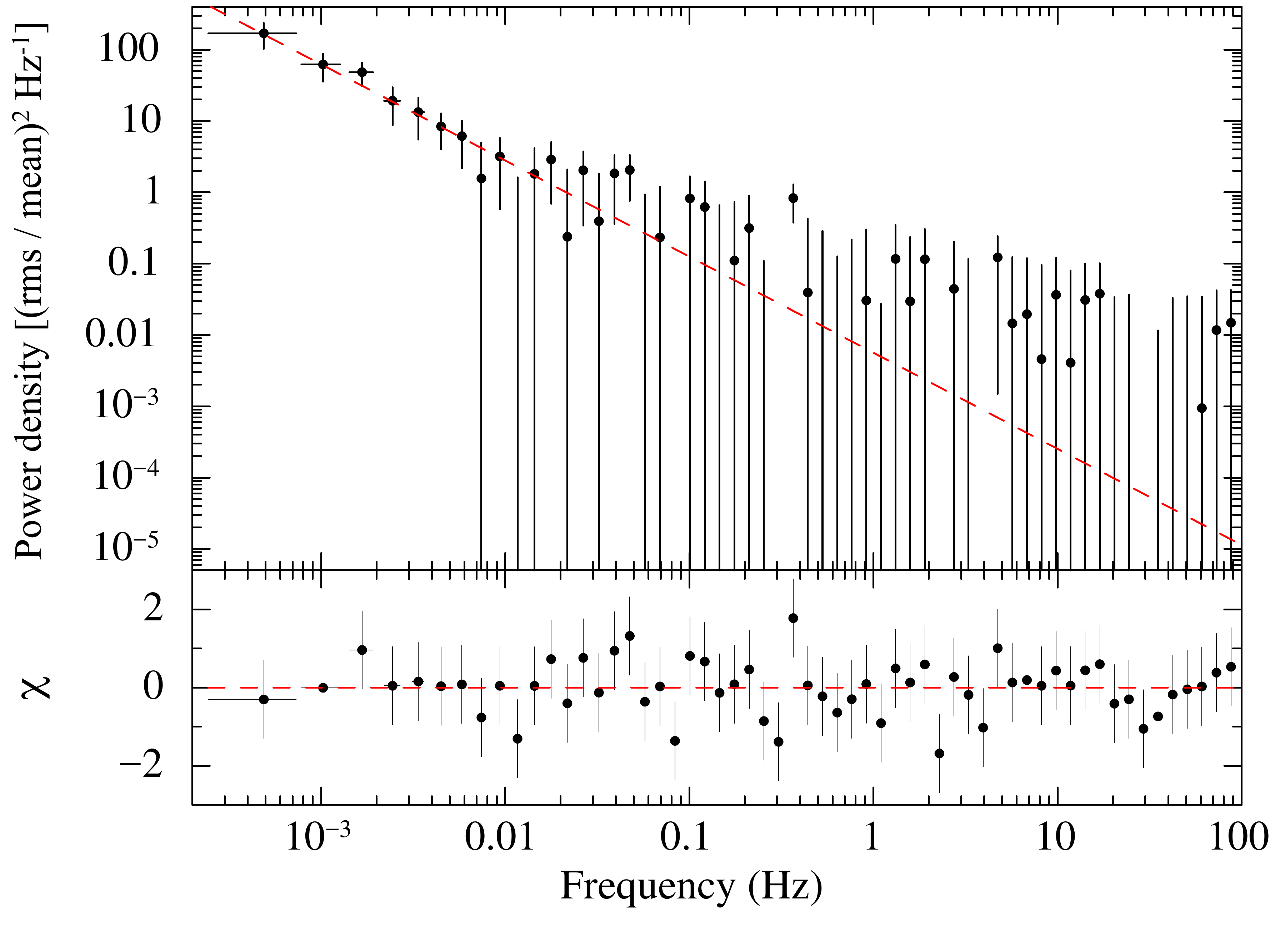}
\vspace{-0.1cm}
\caption{White noise-subtracted power density spectrum of \src\ in the 0.1~mHz -- 100~Hz frequency range, computed from the EPIC pn data filtered over the 0.3--10~keV energy range. The red dashed line represents the best-fitting power law model (see the text for details). Post-fit residuals in units of standard deviations are shown in the {\it bottom panel}.}
\label{fig:psd}
\end{center}
\end{figure}

To study the aperiodic time variability of \src\ more in detail, we extracted the PDS as follows: we first sampled the EPIC pn light curve with a time bin of 2$^{-12}$~s ($\sim 244$~$\mu$s), and calculated power spectra into 2$^{11}$~s (2048~s) long consecutive time intervals (each spectrum is thus characterised by a frequency resolution of $\simeq0.5$~mHz and a Nyquist limiting frequency of 2048~Hz). We then normalised each power spectrum following the prescription of Leahy et al. (1983), averaged the 30 spectra so evaluated, and rebinned the resulting spectrum as a geometric series with a step of 0.2. We estimated a counting statistics (Poisson) noise level of $1.9999(2)$~Hz$^{-1}$ by fitting a constant term to the PDS above 200~Hz, and subtracted it from each power estimate. The white noise-subtracted PDS was then converted to fractional variance per unit frequency (Belloni \& Hasinger 1990; Miyamoto et al. 1991). The stochastic variability of \src\ is characterised by an increasing noise component towards lower frequencies (i.e., red noise), which is adequately described by a power law, $P(\nu) \propto \nu^{-\beta}$, with $\beta=1.3\pm0.1$ ($\chi^2_{{\rm red}}= 1.0$ for 58 d.o.f.; see Fig.~\ref{fig:psd}). We obtained an identical slope when averaging power spectra computed over shorter time lengths (down to 512~s), and/or when applying different geometrical rebinning factors to the averaged spectrum. We did not find any evidence for a turnover in the noise component at low frequencies, not even in the power spectrum extracted using the whole duration of the time series (i.e., down to $T_{\rm obs}^{-1}\simeq20$~$\mu$Hz). We integrated the best-fitting model for the PDS over the 0.5~mHz --100~Hz frequency interval, and estimated a fractional rms variability amplitude of $\sim47$\% for the noise component over this frequency band ($\sim39$\% over the 0.5--10~mHz band). We also extracted the PDSs separately over the 0.3--2~keV and 2--10~keV energy bands, in the high and low modes, as well as over the first 49~ks and the last 16~ks of the observation (the flaring episodes were excluded from the time series). We did not find any change of the functional shape of the PDS either as a function of energy, flux or time. We repeated the analysis using the \nustar\ FPMs data, accounting for the 19 gaps in the time series due to Earth occultation and passages of the satellite through the SAA. 
We chose to compute the PDS from the co-added FPM light curves rather than to adopt the real component of the cross spectrum from the two light curves (a.k.a. co-spectrum), as the spurious effects resulting from the instrumental average dead time of $\simeq2.5$~ms are expected to be negligible at the low average count rate of \src\ ($\simeq0.06$~counts~s$^{-1}$). Moreover, the PDS provides a better sensitivity to the presence of possible variability features than the co-spectrum, as the uncertainties on the powers computed with the PDS are smaller than those obtained with the co-spectrum by a factor of $\simeq2^{1/2}$ (see Bachetti et al. 2015). We derived a best-fitting power law index fully consistent with that inferred from the analysis of the EPIC pn data. On the other hand, a similar analysis could not be performed using the \cxo\ and \swift\ XRT data owing to the coarser time resolution of the ACIS-I and the XRT, as well as the scarce number of source photon counts available.

\begin{figure*}
\centering
\includegraphics[width=1\textwidth]{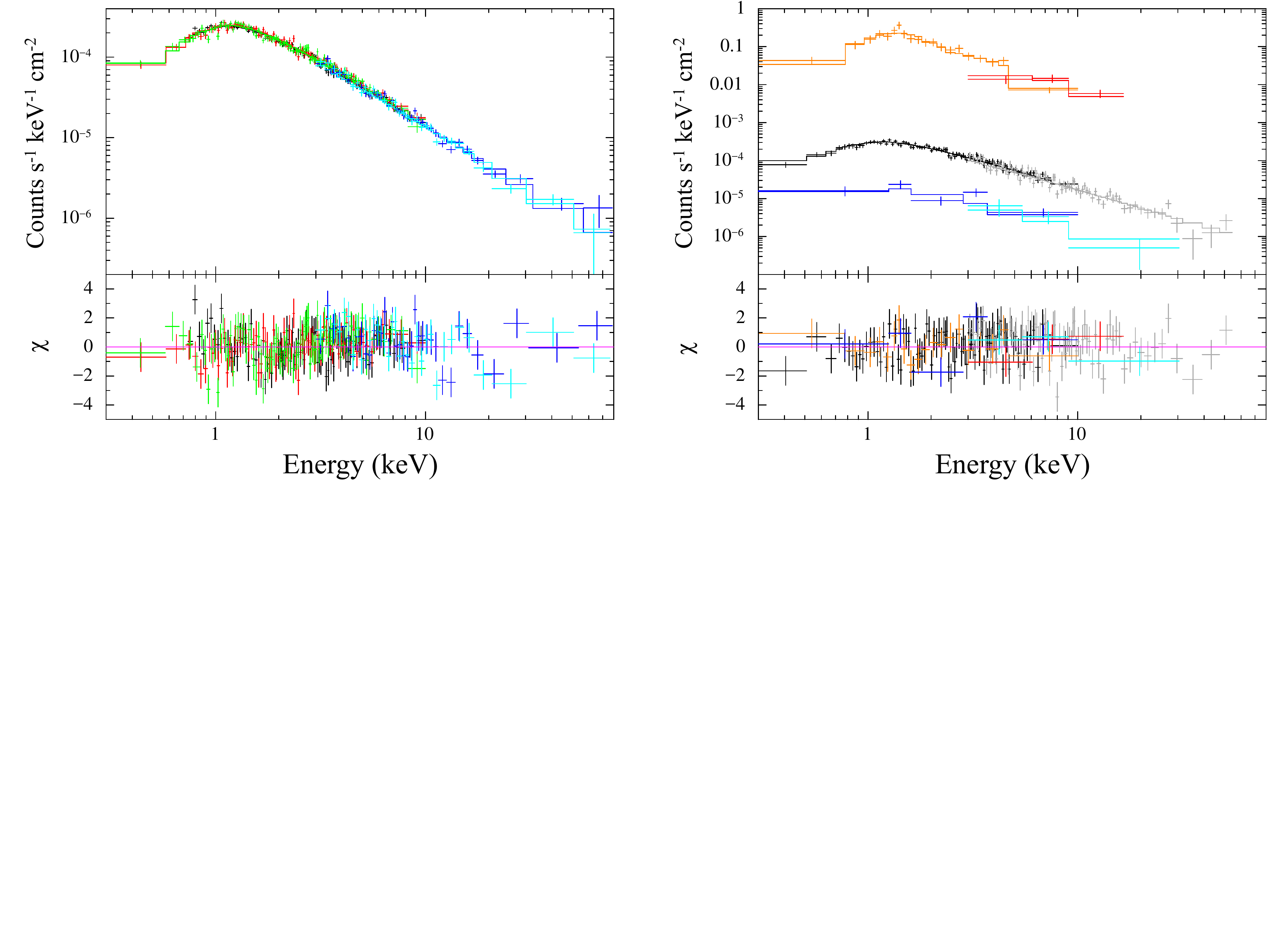}
\vspace{-7cm}
\caption{Background-subtracted X-ray spectra of \src\ extracted over the 0.3--79~keV energy range from the simultaneous \xmm\ and \nustar\ observations on 2018 June 20--21. {\it Left-hand panel}: spectra relative to the average emission. Black refers to the pn, red to MOS\,1, green to MOS\,2, blue to FPMA and cyan to FPMB. {\it Right-hand panel}: spectra relative to the flare, high and low X-ray modes separately. Only the \xmm\ EPIC MOS1 (0.3--10~keV) and \nustar\ FPMA (3--79~keV) spectra are shown for plotting purposes. Orange and red refer to the flare mode, black and grey to the high mode, and blue and cyan to the low mode. In both panels the best-fitting absorbed power law model is indicated by the solid lines, and post-fit residuals in units of standard deviations are shown in the {\it bottom} panel. Data were re-binned to better visualise the trend in the spectral residuals. See the text and Table~\ref{tab:statelog} for details.
}
\label{fig:xmm_nustar_spectra}
\end{figure*}

\subsubsection{Spectral properties}
\label{sec:xrayspectral}

We performed all spectral fits within the \texttt{Xspec} spectral package (v.~12.10.1; Arnaud 1996). We modelled the photoelectric absorption by the interstellar medium along the line of sight via the Tuebingen-Boulder model (\texttt{TBabs} in \texttt{Xspec}), adopting the photoionization cross-sections from Verner et al. (1996) and the chemical abundances from Wilms et al. (2000). We fitted an absorbed power law model to the average \xmm\ and \nustar\ spectra, including a renormalization factor to account for intercalibration uncertainties across the different instruments (these factors were always found consistent with each other within the uncertainties in all subsequent fits). We obtained a satisfactory description of the spectra, with $\chi^2_{{\rm red}} = 1.10$ for 707 d.o.f. The best-fitting parameters were power law photon index of $\Gamma=1.64\pm0.01$ and absorption column density of $N_{\rm H}=(5.3\pm0.1)\times 10^{21}$~cm$^{-2}$. The unabsorbed flux was $F_X=(1.13\pm0.02) \times 10^{-11}$ \flux\ over the 0.3--79~keV energy band, translating into a luminosity of $L_X=(2.16\pm0.04) \times 10^{34}~d_4^2$ \lum\ over the same band assuming isotropic emission. A thermal bremsstrahlung model was instead rejected by the data ($\chi^2_{{\rm red}} = 1.29$ for 707 d.o.f.). More physically-motivated models for the continuum emission (e.g. Comptonization models) are beyond the scope of this paper.

The spectra together with the best-fitting absorbed power law model and the post-fit residuals are shown in the left-hand panel of Fig.~\ref{fig:xmm_nustar_spectra}. We did not observe structured residuals over restricted energy intervals across the whole energy range that might hint at the presence of narrow and/or slightly broadened emission/absorption features. We set upper limits on the emission associated with the iron K$\alpha$ complex by superimposing to the power law component Gaussian models centred either at 6.4~keV (corresponding to the neutral or low-ionised Fe~I~K$\alpha_1$ and K$\alpha_2$ atomic transitions), 6.7~keV (Helium-like Fe~XXV), or 6.97~keV (Hydrogen-like Fe~XXVI). In all fits, the absorption column density and the power law photon index were fixed to their best-fitting values, whereas the Gaussian width was fixed to zero to mimick lines narrower than the intrinsic spectral energy resolution of the different instruments. We derived 3$\sigma$ upper limits of 63, 57 and 73~eV for the equivalent width of any Gaussian-like feature at 6.4, 6.7 and 6.97~keV, respectively.

To check for possible spectral variations either as a function of time, energy and/or count rate without making any assumption on the spectral shape, we first extracted energy vs. time images by binning the source counts collected by EPIC pn into 80 s-long time bins and 50 eV-wide energy channels, and those collected by \nustar\ FPMs into 2.3 ks-long time bins and 1 keV-wide energy channels. We then normalised these images to the time-averaged energy spectrum and the energy-integrated light curve. Visual inspection of the normalised images did not yield any evidence for the presence of structured excesses or deficits of counts with respect to the time-averaged spectrum, i.e. for the presence of emission or absorption features in the spectrum that vary in time and/or energy. We also tried different binnings, and found the same result. As a second step, we computed a few hardness ratios (HRs) between the time series filtered in the same energy intervals used to compute the cross-correlation functions in the X-ray band (see Sect.~\ref{sec:xraytiming}). We adopted different time bins between 30 and 60~s (40 and 80~s) for EPIC (FPMs), so as to guarantee sufficient photon counting statistics in the bins corresponding to the low mode, the mode switchings and the shortest flaring episodes. We also evaluated the HRs by employing an adaptive rebinning of the source light curves (see e.g. Bozzo et al. 2013), so as to achieve a $S/N\geq 10$ in each soft time bin. In none of these cases we did find conspicuous variation of the HR. We repeated these analyses using slightly different selections for the energy ranges, and found the same result.

\begin{table*}
\begin{center}
\caption{Journal of the X-ray modes of \src\ from the simultaneous \xmm\ EPIC and \nustar\ data sets, and results of the spectral fits.}
\label{tab:statelog}
\resizebox{2.04\columnwidth}{!}{
\begin{tabular}{lcccccccc}
\hline \hline 
Mode                                    	& Instrument    		& Net exposure& Net count rate  	& $\Gamma$ 					& Absorbed flux 				& Unabsorbed flux 					& Luminosity 						& $\chi^2_{{\rm red}}$ (d.o.f.)		\\
                                                	&                      		& (ks)                & (counts s$^{-1}$) 	&							& (10$^{-12}$ \flux)				& (10$^{-12}$ \flux)					& (10$^{34}~d_4^2$ \lum)				&					\\
\hline
\multirow{5}{*}{\textit{Flare}}      & EPIC pn                 & 1.0          	& $1.54 \pm 0.04$  	& \multirow{5}{*}{$1.68\pm0.04$} 	& \multirow{5}{*}{$10.7\pm0.5$}		& \multirow{5}{*}{$27\pm2$}			& \multirow{5}{*}{$5.2\pm0.4$}	 		& \multirow{5}{*}{0.89 (75)}		 \\ 
                                                  & EPIC MOS\,1         & 1.0          	& $0.44 \pm 0.02$  	     \\
                                                  & EPIC MOS\,2         & 1.0          	& $0.51 \pm 0.02$   	    \\      
                                                  & FPMA          		& 0.4     		& $0.13 \pm 0.02$  	       \\
                                                  & FPMB          		& 0.4         	& $0.13 \pm 0.02$  	       \\
\hline                                                                                                                                                                      
\multirow{5}{*}{\textit{High}}       & EPIC pn                 & 31.0          	& $0.728 \pm 0.006$ & \multirow{5}{*}{$1.61\pm0.01$} 	& \multirow{5}{*}{$5.42\pm0.06$}	& \multirow{5}{*}{$14.2\pm0.3$}	 		& \multirow{5}{*}{$2.72\pm0.06$}	 	& \multirow{5}{*}{1.19 (437)}		 \\ 
                                                  & EPIC MOS\,1         & 31.3          	& $0.216 \pm 0.003$       \\
                                                  & EPIC MOS\,2         & 30.8          	& $0.248 \pm 0.003$      \\      
                                                  & FPMA          		& 19.9     		& $0.081 \pm 0.002$         \\
                                                  & FPMB          		& 20.0         	& $0.073 \pm 0.002$         \\
\hline                          
\multirow{5}{*}{\textit{Low}}       & EPIC pn                  & 5.7       		& $0.063 \pm 0.008$ & \multirow{5}{*}{$1.5\pm0.1$} 	& \multirow{5}{*}{$0.56\pm0.09$}		& \multirow{5}{*}{$1.6^{+0.5}_{-0.4}$}	& \multirow{5}{*}{$0.31^{+0.10}_{-0.08}$}	 & \multirow{5}{*}{1.05 (26)}		 \\   
                                                 & EPIC MOS\,1          & 5.8       		& $0.016 \pm 0.002$       \\
                                                 & EPIC MOS\,2          & 5.7       		& $0.018 \pm 0.002$       \\      
                                                 & FPMA           		& 4.2         	& $0.007 \pm 0.002$   \\
                                                 & FPMB           		& 4.2           	& $0.005 \pm 0.002$   \\
\hline                                                                                  
\end{tabular}
}
\end{center}
{\bf Notes.} The net count rates refer to the 0.3--10~keV energy band for \xmm\ data and to the 3--79~keV energy range for \nustar\ data. Spectra were fitted with an absorbed power law model with column density fixed to the average value, $N_{\rm H}=5.3\times 10^{21}$~cm$^{-2}$. Fluxes and luminosities are quoted over the 0.3--79~keV energy band. Luminosities were computed assuming a distance of $D=4$~kpc and isotropic emission.
\end{table*}

Nevertheless, a comprehensive analysis of several \xmm\ data sets of the tMSP \object{\psr} over the past 5 years revealed a non-negligible contribution from a soft, thermal component during the X-ray high mode, which was modelled in terms of radiation from both the NS polar caps and the accretion disk (with fractional contributions of $\sim7$ and $\sim3$\%, respectively; Campana et al. 2016; Coti Zelati et al. 2018). Hence, we extracted the \xmm\ EPIC and \nustar\ spectra separately for each mode, and carried out a count rate-resolved spectral analysis. We fitted an absorbed power law model to the spectra, fixing $N_{\rm H}$ at the best-fitting average value and allowing the photon index to vary (as for the average spectra, renormalization factors were included to account for intercalibration uncertainties, and turned out to be always consistent with each other within the uncertainties). In all cases we obtained an adequate description of the data, with no need for an additional thermal component -- either a blackbody model, a model describing the emission from a NS hydrogen atmosphere\footnote{The NS mass, radius and distance were held fixed to $M_{{\rm NS}} = 1.4~M_{\sun}$, $R_{{\rm NS}} = 10$~km and $D=4$~kpc.} (\texttt{nsatmos}; Heinke et al. 2006), or models reproducing the multicolour blackbody emission from an accretion disk (\texttt{diskbb}, \texttt{diskpbb} or \texttt{diskfbb}) -- in any of the modes. 

\begin{figure}
\begin{center}
\includegraphics[width=0.47\textwidth]{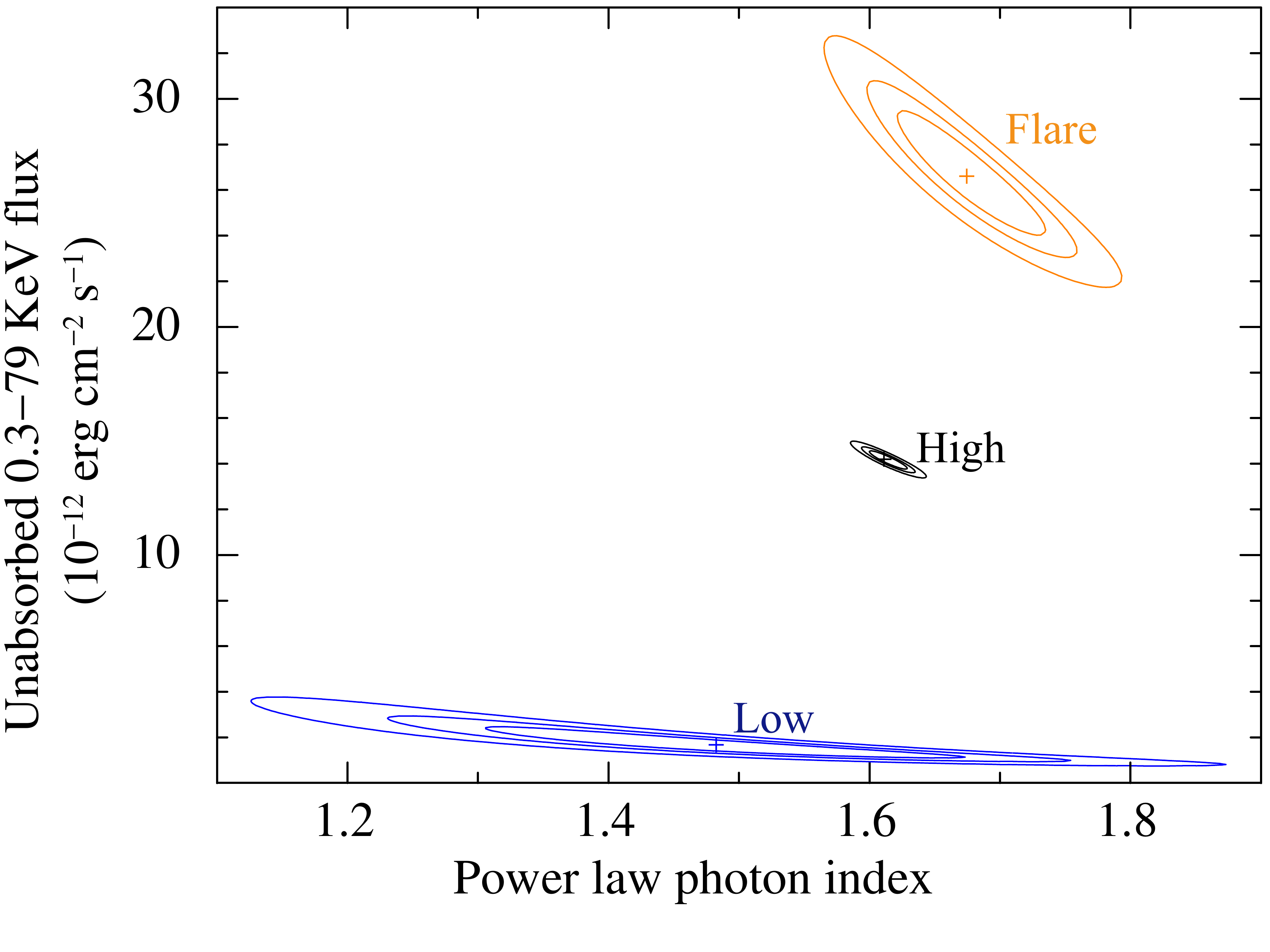}
\vspace{-0.1cm}
\caption{1, 2 and 3$\sigma$ confidence regions for the power law photon index as a function of the unabsorbed flux over the 0.3--79~keV energy band for the best-fitting models of the flare (orange), high (black) and low (blue) mode spectra of \src. The crosses mark the best-fitting values (see Table~\ref{tab:statelog} for details).}
\label{fig:contours}
\end{center}
\end{figure}

The results of the spectral analysis for the three modes are summarised in Table~\ref{tab:statelog}, whereas the spectra, best-fitting models and post-fit residuals are shown in the right-hand panel of Fig.~\ref{fig:xmm_nustar_spectra}. The 1, 2 and 3$\sigma$ contour plots for the power law index and the unabsorbed broad-band flux derived from the spectral fits of each X-ray mode (corresponding to $\Delta\chi^2$ = 2.30, 4.61 and 9.21, respectively) are reported in Fig.~\ref{fig:contours}. They show that the values for the photon index in the different modes are compatible with each other (and with the average value) within 2$\sigma$. We also fitted an absorbed power law model to the broad-band X-ray spectra extracted separately for each flaring event and for each of the three longest episodes of low mode. We resorted to the Cash statistics owing to the paucity of counts available, and evaluated the quality of each fit by Monte Carlo simulations using the \texttt{goodness} command within \texttt{Xspec}. The power law photon indexes and fluxes turned out to be compatible with each other and with the values derived for the corresponding flare and low modes within the large uncertainties.

Finally, we fitted an absorbed power law model to the spectrum extracted using the last 16~ks of the \xmm\ observation and after removal of the time intervals corresponding to the flaring events. We obtained a statistically acceptable result ($\chi^2_{{\rm red}} = 0.96$ for 260 d.o.f.). The best-fitting photon index was $\Gamma=1.58\pm0.02$, again compatible within the uncertainties with that measured during both the high and low modes. The unabsorbed flux was $(8.5\pm0.4) \times 10^{-12}$ \flux\ over the 0.3--79~keV energy range (yielding a luminosity $[1.62\pm0.08] \times 10^{34}~d_4^2$ \lum\ in the same band), i.e. a factor of $\sim1.7$ lower than that maintained in the high mode during the first part of the \xmm\ observation. We conclude that there are no significant changes in the X-ray spectral shape as a function of the count rate and/or time.

\subsubsection{The long-term light curves}
\label{sec:xraylongterm}

\begin{figure}
\begin{center}
\includegraphics[width=0.94\textwidth]{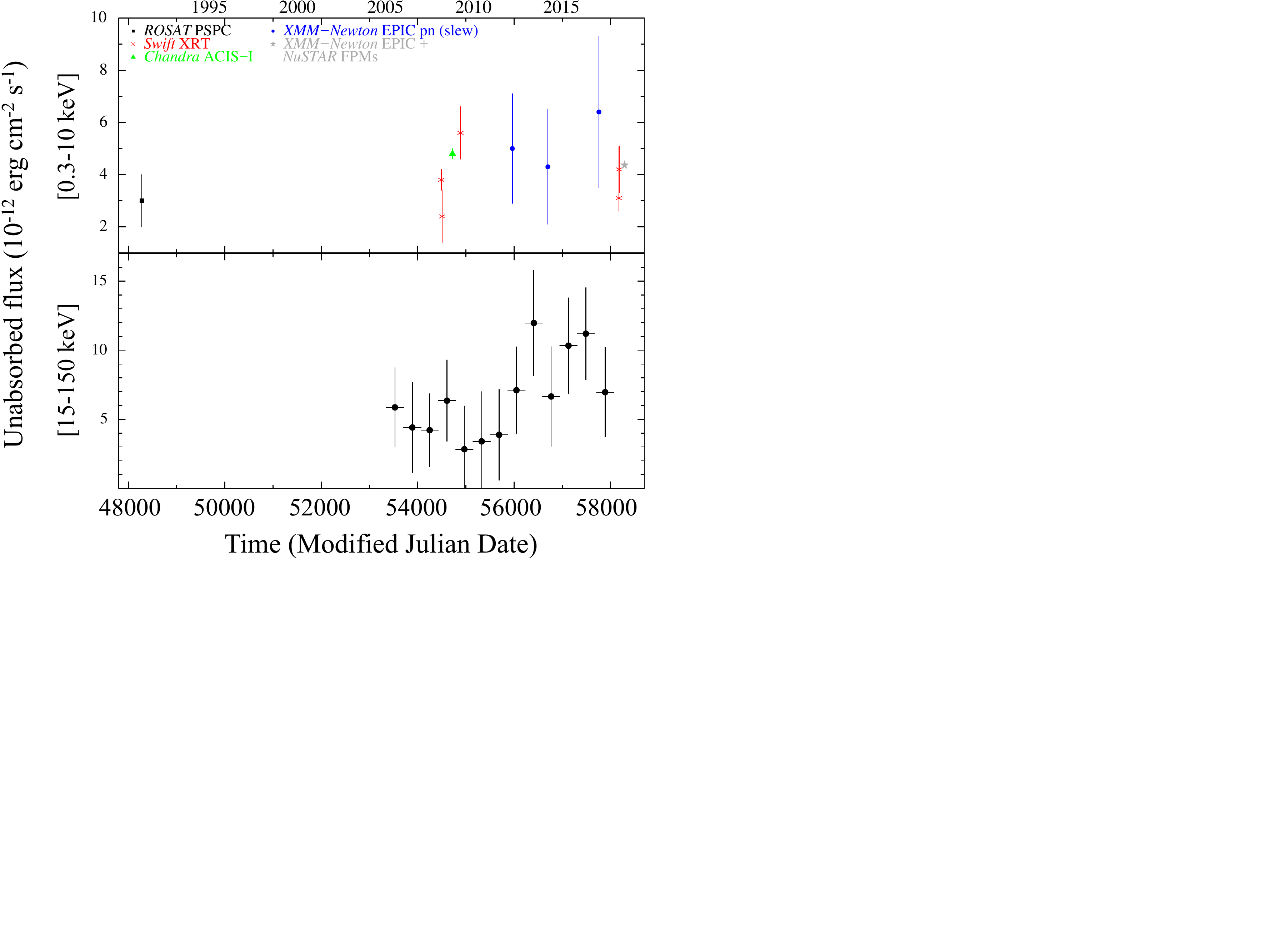}
\vspace{-5.5cm}
\caption{Long-term light curves of \src\ in the X-ray bands. {\it Top panel}: over the 0.3--10~keV energy range, obtained from data with different observatories in the time interval from 1991 January 18 to 2018 June 21. {\it Bottom panel}: over the 15--150~keV energy range, obtained from data with the \swift\ BAT in the time interval from 2004 December 8 to 2017 October 1 with a time bin of 360~d.}
\label{fig:lc_long}
\label{fig:bat_lcurve}
\label{fig:fermi_lcurve}
\end{center}
\end{figure}

The archival observations of the field allowed us to extract the long-term light curve of \src\ in the soft X-ray band over an almost three-decade-long time span, between 1991 and 2018. We first fit an absorbed power law model to all \swift\ XRT and \cxo\ ACIS-I spectra. The absorption column density was fixed at  $N_{\rm H}=5.3\times 10^{21}$~cm$^{-2}$, the best-fitting value inferred from the joint fit of the \xmm\ and \nustar\ spectra. The power law photon index was allowed to vary, and turned out to be consistent with being the same in all cases within the uncertainties. This implies no remarkable changes of the source spectral shape over the time span covered by the observations. On the other hand, the paucity of source counts collected during the \ros\ and \xmm\ slew observations precluded any meaningful spectral modelling of these data sets. Hence, we converted the measured count rates into fluxes using the Portable, Interactive Multi-Mission Simulator (\texttt{PIMMS}, v. 4.9), assuming the best-fitting absorbed power law model derived from the analysis of the \xmm\ and \nustar\ data (see Sect.~\ref{sec:xrayspectral}). The time evolution of the unabsorbed flux of \src\ over the 0.3--10~keV energy band is shown in the top panel of Fig.~\ref{fig:lc_long}. The source has maintained a relatively steady emission level over the time span covered by the observations in the past 10.5~yr, also comparable to that measured in 1991. The slight scatter between the data points, particularly evident for the case of the \swift\ XRT observations, is likely due to the fact that the short exposures of these pointings presumably sampled different modes (or a blend of different modes) at the different epochs. Fitting a constant term to the data points yields $F_{0.3-10 {\rm~keV}}=(4.36\pm0.03) \times 10^{-12}$~\flux\ ($\chi^2_{{\rm red}} = 2.1$ for 10 d.o.f.).

We modelled again the cumulative \integ\ IBIS/ISGRI and \swift\ BAT spectra with an absorbed power law model with parameters fixed to those derived from the joint fit of the \xmm\ and \nustar\ data. We estimated unabsorbed fluxes of $(5.2\pm0.8) \times 10^{-12}$ \flux\ over the 20--50~keV energy range from the IBIS/ISGRI data (over the period 2003--2018; the spectral fit yielded $\chi^2_{{\rm red}} = 0.7$ for 4 d.o.f.) and $(8\pm1) \times 10^{-12}$ \flux\ over the 15--150~keV energy range from the BAT data (between 2004 and 2017; the spectral fit yielded $\chi^2_{{\rm red}} = 1.24$ for 7 d.o.f.). We extracted the hard X-ray light curve of \src\ again performing a count rate-to-flux conversion of the light curve downloaded from the Palermo BAT survey website, assuming the \xmm\ plus \nustar\ spectrum (see the second panel of Fig.~\ref{fig:bat_lcurve}). \src\ has maintained a relatively steady flux in the hard X-ray band since 2004 (fitting a constant to the light curve yielded $\chi^2_{{\rm red}} = 0.77$ for 12 d.o.f.).

\subsubsection{Upper limits on other persistent point sources compatible with FL8Y\,J1109.8$-$6500}
\label{sec:ul}

We employed the \cxo\ datasets to set upper limits on the net count rate for any persistent point-like soft X-ray source located within the error ellipse of FL8Y\,J1109.8$-$6500 other than \src\ (we did not consider the \xmm\ EPIC MOS and \nustar\ imaging data because the adopted instrumental setups provided a much smaller coverage of the error ellipse of the \fermi/LAT source compared to \cxo). We derived an upper limit of $8\times10^{-4}$~counts~s$^{-1}$ over the 0.5--7.0~keV energy band (at the 90\% c.l.), by means of the \texttt{srcflux} tool within \texttt{CIAO}. This limit translates into an unabsorbed flux $F_X<(0.3-1.9) \times 10^{-13}$ \flux\ over the 0.3--10~keV energy band. This value was estimated with \texttt{PIMMS} assuming an absorbed power law spectrum with column density equal to the contribution expected within the Galaxy towards the field ($N_{\rm H}\sim6.7 \times 10^{21}$~cm$^{-2}$; Willingale et al. 2013), and photon index in the range $1<\Gamma<4$. This is a factor $\gtrsim 20$ smaller than the average unabsorbed flux of \src\ within the same energy range (see Sect.~\ref{sec:xraylongterm}).

\subsection{The radio non-detections}
\label{sec:radioul}

\begin{figure}
\begin{center}
\includegraphics[width=1.04\textwidth]{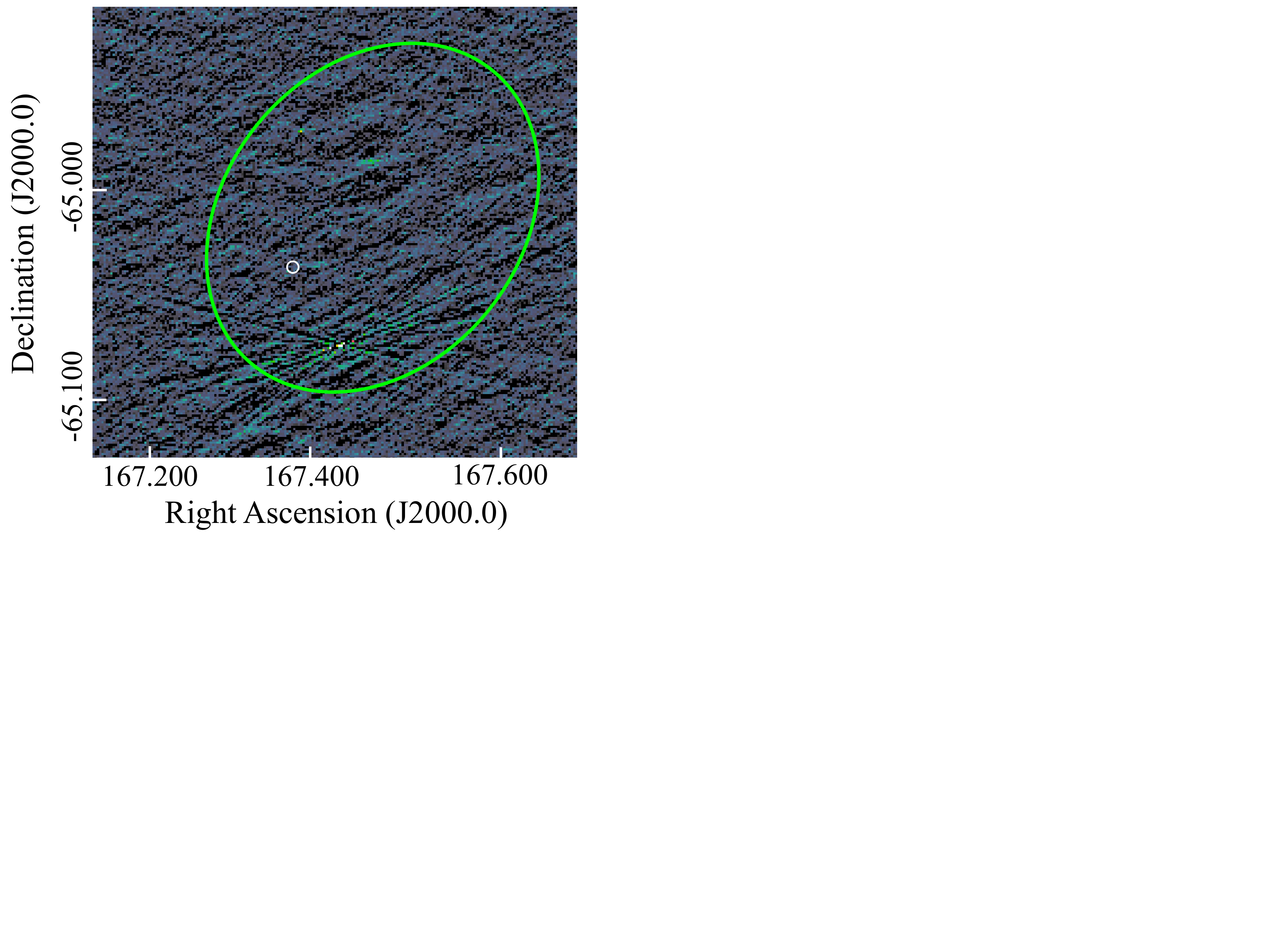}
\vspace{-6.5cm}
\caption{Naturally-weighted 7.25-GHz image of the sky region around \src, extracted from our 4.8 h-long ATCA observation on 2018 June 20. A logarithmic scale was adopted for better visualization. The white circle is centred on the optical position of \src, with an enlarged radius of 10 arcsec for imaging purposes. The green ellipse denotes the error region for the position of the \fermi/LAT gamma-ray source. 
}
\label{fig:fov_atca}
\end{center}
\end{figure}

We did not detect any radio counterpart to \src\ in our ATCA observations. We determined 3$\sigma$ upper-limits on the flux density at the position of \src\ of 27\,$\mu$Jy beam$^{-1}$ at 5.5\,GHz and 24~$\mu$Jy beam$^{-1}$ at 9.0\,GHz. Stacking the two images provided our deepest limit of 18~$\mu$Jy beam$^{-1}$ at a central frequency of 7.25~GHz (see Fig.~\ref{fig:fov_atca} for the stacked image). Assuming a flat radio spectrum as typically observed in accreting compact objects, this value translates into a 5-GHz luminosity of $L_{\rm 5~GHz} = 4\pi \nu F_{5~{\rm GHz}}D^2 < 2.5 \times10^{27}~d_4^2$ \lum. The radio counterpart was also not detected at either frequency or in our stacked image when dissecting the observation into the time intervals corresponding to the X-ray high and low modes recorded with \nustar\ (see Sect.~\ref{sec:xraytiming}). From the stacked images, we measured 3$\sigma$ upper limits of $F_{7.25~{\rm GHz}}<24$~$\mu$Jy beam$^{-1}$ and $F_{7.25~{\rm GHz}}<57$~$\mu$Jy beam$^{-1}$ during the periods of high and low X-ray modes, respectively. Assuming again a flat spectrum, these values translate into luminosities of $L_{5~{\rm GHz}}<4\times10^{27}~d_4^2$~\lum\ and $L_{5~{\rm GHz}}<10^{28}~d_4^2$~\lum, respectively. In Sect.~\ref{sec:phenomcompar} we will use these limits and the luminosities over the 1--10~keV energy interval derived from the analysis of the simultaneous X-ray data (see Sect.~\ref{sec:xrayspectral}) to place \src\ on the X-ray vs. radio luminosity plane for accreting compact objects, comparing the source emission properties with those observed in accreting NSs and, in particular, other tMSPs.

\begin{figure*}
\begin{center}
\includegraphics[width=1.0\textwidth]{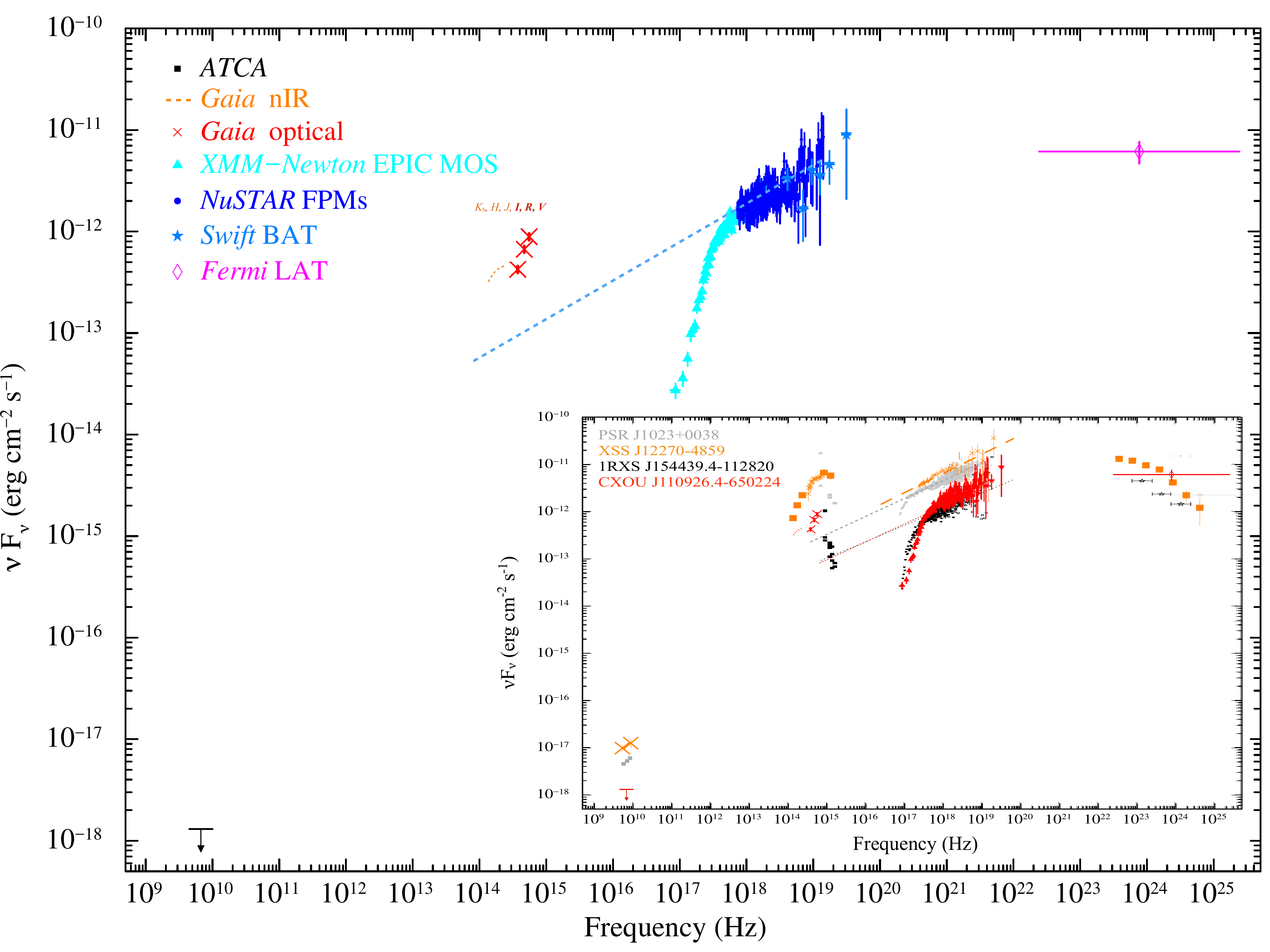}
\caption{{\it Large panel}: time-averaged spectral energy distribution of \src\ from the radio to the gamma-ray band. Vertical bars represent 1$\sigma$ statistical uncertainties, whereas arrowheads indicate 95\% upper limits. The blue dashed line represents the best-fitting power-law model for the average X-ray emission detected with \xmm\ and \nustar, and its extrapolation to the optical and nIR bands. The 
X-ray spectrum has not been corrected for absorption. {\it Inset}: SEDs of \src, \object{\psr} (taken from Bogdanov et al. 2015), \object{\xss} (from de Martino et al. 2015) and \object{1RXS\,J154439.4$-$112820} (from Bogdanov 2016 for the optical, UV and X-ray data) plotted together to highlight their similarities among these systems in the sub-luminous accretion disk state (we note in particular the virtually identical slope of the power law model in the X-ray band).}
\label{fig:sed}
\end{center}
\end{figure*}

\subsection{The spectral energy distribution}
\label{sec:sed}

The extensive multi-wavelength datasets collected for \src\ allow us to extract the time-averaged spectral energy distribution (SED) over a broad range spanning 17 orders of magnitudes in photon frequency, from the radio to the gamma-ray band. The SED is shown in Fig.~\ref{fig:sed}. 

The line-of-sight extinction towards \src\ over the whole {\it Gaia} wavelength bandpass has not been determined (see Andrae et al. 2018). To infer the de-reddened optical and nIR brightness of \src, we substituted the values for its magnitude and its colour quoted in the {\it Gaia} catalogue, $G=20.08$ and $G_{B_{\rm p}}-G_{R_{\rm p}}=1.06\pm0.02$, into the conversion formulas by Evans et al. (2018). This allowed us to derive the magnitudes in the Johnson-Cousins (Landolt 2009) and 2MASS filters: $V\sim20.3$, $R\sim19.8$, $I\sim19.4$, $J\sim18.7$, $H\sim18.3$ and $K_{\rm s}\sim18.2$. 
We did not consider the $B$-band magnitude measured with the \xmm\ OM as representative of the average source brightness, as \src\ was detected only in a couple of exposures (see Sect.~\ref{sec:optprop}). The derived nIR magnitudes should be taken with caution, as these were obtained extrapolating the {\it Gaia} magnitude colour outside the nominal {\it Gaia} bandpass. The expected line-of-sight extinction in the $V$ band can be estimated as $A_V = 1.85 \pm 0.07$, assuming the neutral hydrogen column density derived from the X-ray spectral fits, $\nh = 5.3 \times 10^{21}$~cm$^{-2}$, and the empirical conversion factor $\nh/A_{\rm V}=(2.87 \pm 0.12) \times 10^{21}$~cm$^{-2}$~mag$^{-1}$ (Foight et al. 2016\footnote{The formula was derived from X-ray and optical studies of supernova remnants using the chemical abundances by Wilms et al. (2001), i.e. the same that we assumed to evaluate the absorption column density towards \src.}). This value is smaller than the Galactic integrated extinction in the same band at the source position derived using the extinction maps by Schlafly \& Finkbeiner (2011), $A_V \sim 2.7$. Based on the relations between the extinctions at different wavelengths (Schlafly \& Finkbeiner 2011), we thus estimated extinction coefficients $A_{R} \sim 1.47$, $A_{I} \sim 1.02$, $A_{J} \sim 0.48$, $A_{H} \sim 0.30$ and $A_{K_{\rm s}} \sim 0.20$. The derived de-reddened optical colours $V-R \sim 0.11$, $R-I \sim -0.05$ and $V-I \sim 0.07$ are consistent with a blackbody at a temperature of $T_{{\rm eff}}\sim11\,000$~K.

We attempted to constrain the location of the turnover between the X-ray and gamma-ray power law slopes by fitting a broken power law model to the SED data points over the 3~keV -- 100~GeV energy range, of the form:
\begin{equation}
\nu F_{\nu}=\begin{cases} k\nu^{\beta_1} & \text{for $7.254\times10^{17}{\rm ~Hz}\leq\nu\leq\nu_b$},\\
    				 k \nu^{\beta_2} \nu_b^{(\beta_1-\beta_2)} & \text{for $\nu_b<\nu\leq2.418\times10^{25}{\rm ~Hz}$}.
  \end{cases}
\end{equation}
Here $\nu_b$ is the break frequency. The best-fitting value for the cut-off energy was $E_b = h\nu_b = 49_{-15}^{+19}$~MeV ($\chi^2_{{\rm red}}= 0.96$ for 197 d.o.f.), where $h$ is the Planck's constant.

The SED of \src\ is shown also together with those obtained for \object{\psr}, \object{\xss} and \object{1RXS\,J154439.4$-$112820} in the inset of Fig.~\ref{fig:sed}.

\section{Discussion}
\label{discuss}

The results of our multi-wavelength study of the unidentified X-ray source \src\ can be summarised as follows. 

The optical counterpart was detected at an average magnitude of $\sim20.1$ over the 3300--10\,500~\AA\ wavelength range. Considerable variability was observed on timescales ranging from hundreds to thousands of seconds. In particular, the optical brightness varied by $\sim0.3-0.4$ mag over an elapsed time of less than a ks (see Table~\ref{tab:phot}). Optical spectra obtained at different epochs showed prominent Balmer emission lines with variable profiles. The de-reddened optical colours are consistent with a blackbody temperature of $T_{{\rm eff}}\sim 11000$~K, as typically observed in accreting systems.
The distance estimated from parallax measurements using different assumptions, the proximity of the source to the Galactic plane ($b\simeq-4.3$\deg), and the presence of optical emission lines at a redshift of $z \simeq 0$, point to a location of \src\ within the Milky Way.

The X-ray spectral shape measured over the 0.3--79~keV energy band with our simultaneous \xmm\ and \nustar\ observations was described adequately by an absorbed power law model with photon index of $\Gamma\sim1.6$, with no evidence for any cutoff. The average luminosity at the epoch of these observations was $\sim2.2\times10^{34}~d_4^2$ \lum\ over the same energy band. The X-ray emission showed dramatic short-term variability: \src\ spent about 60\% of its time in a stable high mode, at a luminosity $\sim2.7\times10^{34}~d_4^2$ \lum\ over the 0.3--79~keV energy band, which unpredictably alternated on timescales as short as tens of seconds to a stable low mode for about 10\% of the time, at a luminosity $\sim3\times10^{33}~d_4^2$ \lum\ over the same band.  
Sporadic flaring activity at a luminosity of  $\sim5\times10^{34}~d_4^2$ \lum\  was also observed on top of a more erratic, variable emission. No significant spectral changes were observed across the three modes. No statistically significant periodic signal was found in the power density spectrum. An upper limit of $A<42$\% (3$\sigma$ c.l.) was set on the rms amplitude for any coherent signal with frequency lower than 1.5~kHz emitted by a pulsar in a tight orbit (orbital period of a few hours) during the high mode. The X-ray aperiodic time variability of \src\ was characterised by a red flicker noise component extending down to a few 10$^{-5}$~Hz, which could be well modelled by a power law with index of $\beta\sim1.3$. 
Our recent X-ray observations were complemented by archival X-ray datasets with different space-born satellites (\ros, \swift, \cxo, \integ), covering a broad energy range. These again revealed a distinctive bimodal behaviour in the net count rates, and showed that the source has maintained a relatively steady average flux level over the time span covered by the observations. 

The spatially associated gamma-ray source was detected by \fermi/LAT at an average luminosity of $\sim1.5\times10^{34}~d_4^2$ \lum\ over the 100 MeV--300~GeV energy band between 2008 and 2016, comparable to the average X-ray luminosity of \src. Its integrated spectrum was well described by a power law with photon index of $\Gamma_\gamma\sim2.6$. 

Radio continuum observations with ATCA (which were simultaneous with \nustar) resulted in a non-detection, with a 3$\sigma$ upper limit on the 5-GHz luminosity of $2.5\times10^{27}~d_4^2$~\lum.

The multiband observational campaign that we conducted on \src\ allows us to compare several phenomenological properties of this source with those already observed in the three confirmed tMSPs (\object{IGR\,J18245$-$2452}, \object{\psr} and \object{\xss}) during their peculiar sub-luminous accretion disk state, as well as in a couple of additional tMSP candidates, \object{1RXS\,J154439.4$-$112820} (Bogdanov \& Halpern 2015; Bogdanov 2016) and the eclipsing binary 1SXPS\,J042749.2$-$670434 (Strader et al. 2016)(see Campana \& Di Salvo 2018 for a comprehensive observational review). 
Before addressing the phenomenological analogies between \src\ and the above-mentioned systems in the next section, we point out that two other X-ray sources have been proposed recently as tMSP candidates: XMM\,J083850.4$-$282759 (Rea et al. 2017) and \object{CXOGlb\,J174804.5$-$244641} amid the globular cluster \object{Terzan 5} (Bahramian et al. 2018a).

\subsection*{A phenomenological comparison with confirmed and candidate tMSPs}
\label{sec:phenomcompar}

Optical variability is commonly observed in the sub-luminous accretion disk state of tMSPs on a wide range of timescales, from seconds (e.g. Shahbaz et al. 2018) to weeks (Kennedy et al. 2018; Papitto et al. 2018). While the presence of a stable mode switching (similar to that detected in the X-ray band) is currently debated, tMSPs (and \src\ as well) are certainly observed to flicker and undergo sporadic flaring episodes in the optical with variable duration between tens of seconds and hours. In almost all cases, these flares correlate with the emissions in the X-ray band (see e.g. Jaodand et al. 2016 for a compilation of simultaneous X-ray and optical time series of \object{\psr}), as well as with those in the nIR and ultraviolet bands (see, e.g. de Martino et al. 2010, 2013; Saitou et al. 2011; Patruno et al. 2014; Bogdanov et al. 2015; Bogdanov \& Halpern 2015; Bogdanov 2016; Strader et al. 2016; Shahbaz et al. 2018). Prominent emission lines in the optical spectrum typical of accreting systems were also observed for \object{\psr} (e.g. Coti Zelati et al. 2014; Bogdanov et al. 2015), \object{\xss} (de Martino et al. 2014) and 1SXPS\,J042749.2$-$670434 (Strader et al. 2016). In \src\ as well as in these other systems, the profiles of the Balmer lines are typically double-peaked. This is indicative of motion within an accretion disk around the NS. 

The bimodal distribution in the net X-ray count rates of \src\ is strikingly similar to that observed over a broad X-ray energy range (and also over years) in almost all the known tMSPs in the sub-luminous accretion disk state (e.g. Saitou et al. 2009; de Martino et al. 2010, 2013; Linares 2014; Tendulkar et al. 2014; Bogdanov et al. 2015; Jaodand et al. 2016; Coti Zelati et al. 2018). The rapid timescale of the switches and the fraction of time spent in each of the three modes are also reproducible, not only from source to source, but also across observations of the same target at different epochs. 
We note, however, that such a remarkable similarity among the tMSPs might represent only a fluke, due to the small sample size of tMSPs caught so far in the sub-luminous accretion disk state, as well as to the detection of an `anomalous' 10 h-long low mode in the X-ray emission of \object{IGR\,J18245$-$2452} (Papitto et al 2013; Linares et al. 2014). Although no flares have been reported yet from the tMSP candidates \object{1RXS\,J154439.4$-$112820} and 1SXPS\,J042749.2$-$670434, intense flares with a variety of morphologies and durations have been observed in almost all high-time resolution \xmm\ and \nustar\ observations of \object{\psr} and \object{\xss} (e.g. de Martino et al. 2010, 2013; Jaodand et al. 2016). Interestingly, a common characteristic shared by \src\ and these two tMSPs is that the onset of (brighter) flaring events is often preceded by relatively long time intervals of low mode. For example, both \object{\xss} and \object{\psr} persisted in the low mode for about $\sim20$ min before undergoing a sustained flaring activity during \xmm\ observations in 2011 January and 2015 November, respectively. There is, however, no evidence for any significant correlation between the duration of the low mode and that of the subsequent flares or their peak intensity. Moreover, in \src, \object{\xss} and \object{\psr}, flares are typically interspersed with time intervals of lower flux where the X-ray emission varies erratically, at odds with the flat-bottomed emission observed during the low mode. The highly reproducible pattern of the light curve in different observations and the remarkable steadiness of both the high and low mode intensity levels over timescales of years suggest a negligible impact of the flaring events on the emission properties in the two modes, in tMSPs as well as in \src.

\begin{table*}[h!]
\begin{center}
\caption{Multi-wavelength emission properties of the confirmed and candidate tMSPs in the sub-luminous accretion disk state; other systems properties are also listed.}
\label{tab:tmsps}
\resizebox{2.05\columnwidth}{!}{
\begin{tabular}{lccccccc}
\hline  \hline 
									& \multicolumn{3}{c}{\large{Confirmed}$^a$}								& \multicolumn{4}{c}{\large{Candidates}$^b$} 	 \\	
\cmidrule(lr){2-4} \cmidrule(lr){5-8}
 		               						& \large{J18245}		      	& \large{J1023} 				& \large{J12270}	        & \large{J1544} 			& \large{J0427} 		& \large{J1748}  	& \large{J1109} 	\\			
\hline \vspace{0.25cm}
{\it Optical properties}$^c$					&						&							&					&					&					&				& \\ \vspace{0.05cm}
Magntitude							& --						& $G\sim16.3$					& $V\sim16.1$			& $G\sim18.6$			& $G\sim17.7$			& --				& $G\sim20.1$	\\	\vspace{0.05cm}
Short-term variability						& --						& \checkmark					& \checkmark			& \checkmark			& \checkmark			& --				& \checkmark \\	 \vspace{0.05cm}	
Orbital modulation						& --						& \checkmark					&  \checkmark			&  \checkmark			&  \checkmark			& --				& -- \\	 \vspace{0.05cm} 		
Spectral emission lines 					& --						& \checkmark					& \checkmark			& \checkmark			& \checkmark			& --				& \checkmark 	\\	 \vspace{0.05cm}
Pulsations								& --						& \checkmark (in high, flare modes)	& --					& --					& --					& --				& --  \\	
\hline \vspace{0.25cm}
{\it X-ray properties}$^d$					&						&							&					&					&					&				& \\ \vspace{0.05cm}
$\Gamma_X$ 			 				& $1.44\pm0.05$ 		 	& $1.62\pm0.02$				& $1.70\pm0.02$		& $1.63\pm0.02$		& $1.68_{-0.08}^{+0.09}$	& $1.5\pm0.1$ 		& $1.63\pm0.02$  \\	\vspace{0.05cm}
$L_X$ (10$^{33}$ \lum)					& $11.2\pm0.5$ 			& $5.2\pm0.1$ 					& $12\pm2$ 			& $16.5\pm0.6$		& $6.3\pm0.4$			& $6.5\pm0.3$		& $21.6\pm0.4$ 	\\ \vspace{0.05cm}	
$L_{X, {\rm high}}$ (10$^{33}$ \lum)			& $13.1\pm0.6$			& $7.9\pm0.1$ 					& $13\pm3$ 			& $25.1\pm0.7$ 		& $10.3\pm0.8$		& --				& $27.2\pm0.6$ 	\\ \vspace{0.05cm}
$L_{X, {\rm low}}$ (10$^{33}$ \lum)			& $2.0\pm0.3$				& $0.87\pm0.04$				& $2.0\pm0.4$			& $1.4\pm0.3$ 			& $3.6\pm0.4$			& --				& $3\pm1$	\\ 	
\vspace{0.05cm}
Pulsations								& \checkmark				& \checkmark (in high mode)		& \checkmark (in high mode)& --				& --					& --				& --  \\	\vspace{0.05cm}
Red noise power law slope				& $1.18\pm0.01$			& $1.05\pm0.05$				& $1.39\pm0.06$		& $1.41\pm0.04$		& $1.5\pm0.1$			& --				& $1.3\pm0.1$ \\		
\hline \vspace{0.25cm}
{\it Gamma-ray properties}$^e$				& 						&							&					&					&					&				& \\ 
$\Gamma_\gamma$   					& --	 					& $2.41\pm0.10$ 				& $2.36\pm0.06$  		& $2.48\pm0.05$ 		& $2.48\pm0.06$  		& --	 			& $2.58\pm0.15$ 	\\ \vspace{0.05cm}	
									&						&							&					& {\small($2.53\pm0.05$)}	& {\small($2.39\pm0.06$)}	&				&   \\	
$L_{\gamma}$ (10$^{33}$ \lum)			& -- 						& $12.5\pm0.4$ 				& $21.9\pm0.7$  		& $24.7\pm0.2$ 		& $6.5\pm0.4$			& --				& $15\pm2$	\\      	
\hline \vspace{0.25cm}
{\it Radio properties}$^f$					&						&							&					&					&					&				& \\  \vspace{0.05cm}
$\Gamma_{r}$ 							& $0.39\pm0.06$			& $0.09\pm0.18$				& $-0.1\pm0.1$ 		& --					& --					& --				& -- \\	\vspace{0.05cm}
$L_r$ (10$^{27}$ \lum)					& 112					& 0.97						& 5.4					& --					& --					& --				& $<2.5$ \\	\vspace{0.05cm}
$L_r$ [high X-ray mode] (10$^{27}$ \lum)		& --						& 0.63 						& --					& --					& --					& --				& $<4.0$ \\	\vspace{0.05cm}				
$L_r$ [low X-ray mode] (10$^{27}$ \lum)		& 112					& 1.9							& --					& --					& --					& --				& $<10.0$ \\	
\hline \vspace{0.25cm}
{\it Other properties}$^g$					&						&							&					&					&					&				& \\  \vspace{0.05cm}
NS spin period (ms)						& 3.93					& 1.69						& 1.69				& --					& --					& --				& -- \\ \vspace{0.05cm}
System orbital period (h)					& 11.03					& 4.75						& 6.91				& 5.80				& 8.80				& --				& -- \\  \vspace{0.05cm}
NS spin-down luminosity ($10^{34}$ \lum)	& -- 						& $4.8\pm0.5$					& $8.3\pm1.8$			& --					& --					& --				& -- \\  \vspace{0.05cm}
NS dipolar magnetic field ($10^8$~G)		& --						& 1.9							& 2.3					& --					& --					& --				& -- \\  \vspace{0.05cm} 
System inclination (\deg)					& --						& $46\pm2$					& 46--55				& $7\pm1$			& $78\pm2$			& --				& --	\\  \vspace{0.05cm}
Spectral type of the companion				& --						& Mid-G V 					& G5					& K6--7V				& mid-to-late K			& --				& -- \\ \vspace{0.05cm}
NS mass ($M_{\sun}$)					& --						& $1.65^{+0.19}_{-0.16}$			& --					& --					& $1.86\pm0.11$		& --				& -- \\ \vspace{0.05cm}
Companion mass ($M_{\sun}$)			& $0.20\pm0.01$			& $0.22\pm0.03$ 				& $\ge0.27$			& $\lesssim0.7$		& $0.65\pm0.08$		& --				& -- \\
\hline
\end{tabular}
}
\end{center}
{\bf Notes.}
Values are not reported for those cases where the corresponding counterpart is uncertain, has not been identified yet or was not observed during the sub-luminous accretion disk state, and also where the X-ray modes and the slope of the red noise component have not been clearly singled out. 
\noindent $^{(a)}$ Systems that have been observed to undergo a state transition.
\noindent $^{(b)}$ Systems that have not been observed to undergo a state transition so far.
\noindent $^{(c)}$ Optical magnitudes are not corrected for extinction and are taken from the second {\it Gaia} data release, except for \xss\ which persisted in a rotation-powered state since the satellite launch in 2013 December.
\noindent $^{(d)}$ All X-ray luminosities were re-computed in this work over the 0.3--79~keV energy band adopting the abundances of Wilms et al. (2000) for the absorption model, to allow a direct comparisons among the different systems. We assumed isotropic emission at the following distances: 5.5~kpc for \object{IGR\,J18245$-$2452} (Harris 1996), 1.37~kpc for \object{\psr} (Deller et al. 2012), 1.6~kpc for \object{\xss} (Jennings et al. 2018), 3.8~kpc for \object{1RXS\,J154439.4$-$112820} (Britt et al. 2017), 2.4~kpc for 1SXPS\,J042749.2$-$670434 (Strader et al. 2016), 5.9~kpc for \object{CXOGlb\,J174804.5$-$244641} (Lanzoni et al. 2010) and 4~kpc for \src\ (this work). 
The luminosities of \object{IGR\,J18245$-$2452} and \object{CXOGlb\,J174804.5$-$244641}, for which no observations exist in the hard X-ray band during the sub-luminous accretion disk state, were obtained extrapolating the best-fitting absorbed power law model for the soft X-ray emission and assuming no spectral cut-off up to 79~keV. Uneclipsed luminosities are reported for 1SXPS\,J042749.2$-$670434.
\noindent $^{(e)}$ The value for the simple power law model is reported for both \object{\psr} and \object{\xss}, although evidence for an exponential cut-off was found in both systems (Torres et al. 2017; Xing et al. 2018). The value reported on the \fermi/LAT eight-year point source list is also quoted in parentheses. Uncertainties are statistical only.  Gamma-ray luminosities are evaluated over the 0.1--300~GeV energy band and for the same distances as in $(d)$. Uncertainties are statistical only. 
$^{(f)}$ Radio luminosities are evaluated at a frequency of 5~GHz and for the same distances as in $(d)$. Upper limits are evaluated assuming a flat spectrum.
\noindent $^{(g)}$ See also Strader et al. (2019).
\end{table*}

\begin{figure*}
\begin{center}
\includegraphics[width=1.0\textwidth]{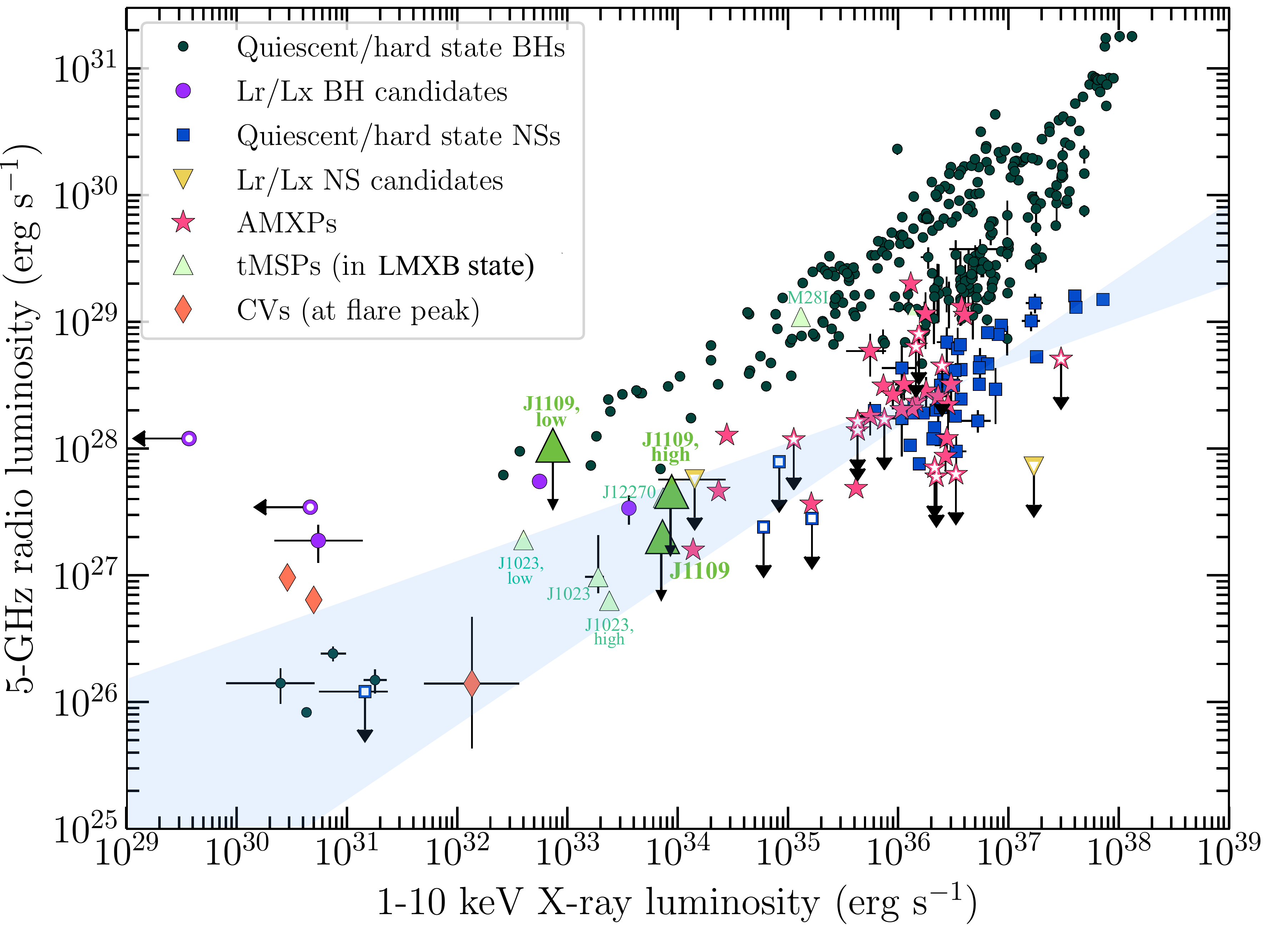}
\vspace{-0.5cm}
\caption{Radio vs. X-ray luminosity plane for different classes of accreting compact objects, adapted from Bahramian et al. (2018b). Data points for \object{\psr} and \src\ (see the green triangles) are plotted also separately for the high and low X-ray modes (Bogdanov et al. 2018 and this work). The data point for \object{IGR\,J18245$-$2452} (M28I) was derived from strictly simultaneous ATCA and \swift\ XRT observations (Ferrigno et al. 2014), and we computed its position on the plane after performing time-resolved spectral analysis of \swift\ XRT data (obs. ID 00032785003). Upper limits are denoted by arrowheads. The cyan shaded area encloses the 3$\sigma$ confidence interval on the shape of the correlation holding for accreting NSs (see Gallo et al. 2018 and the text for more details).}
\label{fig:lx_lradio}
\end{center}
\end{figure*}

The upper limit that we derived on the pulse fractional amplitude for any coherent signal during the high mode, $A<42$\%, does not constrain significantly the presence of X-ray pulsations from \src\ as it is much larger than the typical amplitudes measured for tMSPs in the same mode, $\simeq8$\% (Archibald et al. 2015; Papitto et al. 2015). The aperiodic time variability of \src\ lacks the typical flat-topped low-frequency noise observed in several accreting MSPs (with characteristic frequency within the 0.1--1~Hz interval; see e.g. van Straaten et al. 2005). This appears, however, to be a hallmark of tMSPs. In particular, the functional shape of the noise of \src\ -- a power law with index $\beta \sim 1.3$ -- is similar to that already reported for other members of the class (Papitto et al. 2013, Ferrigno et al. 2014 for the case of \object{IGR\,J18245$-$2452}; Tendulkar et al. 2014 for \object{\psr}; Hern\'andez Santisteban 2016, Papitto et al. 2018b for analogous studies of \object{\psr} in the UV and optical bands, respectively). We also performed a preliminary study of the aperiodic time variability of both \object{1RXS\,J154439.4$-$112820} and 1SXPS\,J042749.2$-$670434 using \nustar\ and \xmm\ data publicly available, and adopting a similar procedure to that described in Sect.~\ref{sec:xraytiming}. We derived again a similar slope, $\beta \sim 1.4-1.5$. 

The X-ray spectral shape of \src\ is virtually identical to that observed for the average emission of all tMSPs in the sub-luminous accretion disk state (e.g. de Martino et al. 2013; Bogdanov 2016; Strader et al. 2016; Coti Zelati et al. 2018). \src\ was always caught at a similar X-ray flux across a broad energy band over the past $\gtrsim15$~yr. The detection during an archival \ros\ observation in 1991 at comparable flux is probably indicative of an even longer X-ray active state, although the lack of sensitive X-ray observations of the field between 1991 and 2003 cannot rule out -- at least in principle -- a possible switch back and forth to a fainter X-ray state over this time period. Similar prolonged ($\gtrsim 8$~yr) sub-luminous states were observed also in the two tMSP candidates \object{1RXS\,J154439.4$-$112820} (Bogdanov 2015) and 1SXPS\,J042749.2$-$670434 (Strader et al. 2016). Furthermore, the case of \src\ is very similar to that of \object{\xss}, which was also detected by \ros, the {\it Rossi X-ray Timing Explorer}, \integ\ and \swift\ BAT before switching to a X-ray fainter disk-free, rotation-powered state in 2013 (e.g. de Martino et al. 2010; Papitto et al. 2014). We note that no pulsar spatially coincident with \src\ is listed in the most updated online version of the Australia Telescope National Facility (ATNF) pulsar catalogue (v.~1.59; see Manchester et al. 2005). However, no deep searches for radio pulses from the source position have been performed to our knowledge. Dedicated studies are required to investigate whether a radio pulsar is turned on during the current state of \src. The dispersion measure expected at the position and distance of \src\ is comparatively large, $DM \approx 390$~cm$^{-3}$~pc according to the state of the art model for the distribution of free electrons in the Galaxy (Yao et al. 2017)\footnote{\texttt{http://119.78.162.254/dmodel/index.php}}.

The gamma-ray emission properties of the \fermi\ source associated to \src\ also closely resemble those observed for \object{\psr} and \object{\xss} and the candidates \object{1RXS\,J154439.4$-$112820} and 1SXPS\,J042749.2$-$670434 in the sub-luminous accretion disk state, i.e. the only known NS LMXBs hitherto known for which a GeV counterpart has been established unambiguously. All these systems show a power-law like spectrum (Torres et al. 2017 and references therein) extending up to a much higher energy compared to the rotation-powered state, and slightly softer ($\Gamma_{\gamma}\sim 2.4-2.5$) than that measured in the X-ray band ($\Gamma \sim 1.5-1.7$). The larger photon counting statistics available for the two brightest gamma-ray sources, \object{\psr} and \object{\xss}, led additionally to the detection of a statistically significant spectral cut-off in these systems, at energy $E_{\rm C} \sim 3$~GeV and $\sim 11$~GeV, respectively (and a slightly harder photon index of $\Gamma_{\gamma}\sim 2.0$, 2.3 compared to the simple power law model; see Torres et al. 2017; Xing et al. 2018). The ratio between the average gamma-ray and X-ray luminosities is also compatible across all these systems when in the sub-luminous accretion disk state, and of the order of unity: $L_{\rm 0.1-300~GeV} / L_{\rm 0.3-79~keV} \sim 0.7$, 2.4, 1.8, 1.5 and 1 for \src, \object{\psr}, \object{\xss}, \object{1RXS\,J154439.4$-$112820} and 1SXPS\,J042749.2$-$670434, respectively. A summary of the multiband emission properties of confirmed and candidate tMSPs in the sub-luminous accretion disk state is reported in Table~\ref{tab:tmsps}.

Notably, the SED reported in Fig.~\ref{fig:sed} indicates that the extrapolation of the best-fitting power-law model for the X-ray spectrum (corrected for the effects of interstellar absorption) towards lower photon frequencies predicts an optical and nIR flux which is smaller than that inferred using {\it Gaia} measurements. This implies that the emission process responsible for the X-ray flux cannot account for the bulk of the optical and nIR emissions. Such optical/nIR excess was reported also for the cases of \object{\xss} (de Martino et al. 2013, 2015), \object{\psr} (Coti Zelati et al. 2014; Bogdanov et al. 2015) and the tMSP candidate \object{1RXS\,J154439.4$-$112820} (Bogdanov 2016), and it was ascribed to emission from the donor and the accretion disk surrounding the NS (see the inset of Fig.~\ref{fig:sed}). The energy of the spectral turnover between the X-ray and gamma-ray power law slopes in \src, although only poorly constrained, is of the same order of magnitude of that estimated for other tMSPs (see, e.g. Tendulkar et al. 2014). X-/gamma-ray observations with instruments covering the MeV--GeV energy range, such as those of the originally proposed mission {\it e-ASTROGAM} (de Angelis et al. 2018), could constrain better the energy of the spectral turnover of \src\ and tMSPs in general. 

The only property that makes \src\ seemingly at odds with the other tMSPs is the non-detection of its radio counterpart, as all tMSPs studied so far were observed to emit in the radio band in the sub-luminous accretion disk state. In these other systems, the radio emission is characterised by a flat to slightly inverted spectrum ($\alpha \sim 0-0.5$, where $F_\nu \propto \nu^{\alpha}$) that is significantly different from the typical pulsar-like steep spectrum ($\alpha < -1.5$) observed during the disk-free, rotation-powered state. The emission is also rapidly variable, a factor of $\sim2-3$ on timescales of minutes (Ferrigno et al. 2014; de Martino et al. 2015; Deller et al. 2015). Radio counterparts were detected also for \object{1RXS\,J154439.4$-$112820} and 1SXPS\,J042749.2$-$670434 at 5~GHz in ATCA observations in 2012 September (Petrov et al. 2013). The radio:X-ray luminosity plane for accreting compact objects including \src\ is shown in Fig.~\ref{fig:lx_lradio} (see Bahramian et al. 2018b for an updated compilation). Recently, Gallo et al. (2018) performed a comprehensive regression analysis aimed at characterising empirically the shape of the correlation between the radio and X-ray luminosities for different classes of accreting compact objects. In particular, they found that the correlation holding for a sample of 41 NSs (including all known accreting X-ray MSPs and the 3 tMSPs) can be described by the following formula:

\begin{dmath}
\label{eq:corr}
{\rm Log}\Bigg(\frac{L_{{\rm 5~GHz}}}{\lum}\Bigg) = (28.40\pm0.15) + 0.44_{-0.13}^{+0.15} \Bigg[{\rm Log}\Bigg(\frac{L_{{\rm 1-10~KeV}}}{\lum}\Bigg) -36.30 \Bigg],
\end{dmath}
where the uncertainties on the constant and the slope are quoted at the 3$\sigma$ c.l. (see the cyan shaded area in Fig.~\ref{fig:lx_lradio}). Assuming that \src\ follows the correlation described by Eq.~(\ref{eq:corr}), we would expect an average radio flux density within the range $\approx8 - 50$~$\mu$Jy at 5~GHz based on the X-ray luminosity estimated from our \xmm\ and \nustar\ observations. Therefore, the 3$\sigma$ flux upper limit of 18~$\mu$Jy beam$^{-1}$ (at 7.25~GHz) that we derived in Sect.~\ref{sec:radioul} would be consistent with the presence of a faint radio counterpart. Furthermore, we note that the shape of the correlation has become more uncertain very recently following the discovery of radio emission from the Be/X-ray binary system \object{Swift\,J0243.6$+$6124} during its outburst decay between 2017 October and 2018 January (van den Eijnden et al. 2018), as well as of bright radio emission from the accreting MSP \object{IGR\,J17591$-$2342} during the early stages of its outburst in 2018 August (Russell et al. 2018). Accounting for these detections leads to an even wider parameter space allowed for accreting NSs on the radio:X-ray luminosity plane (see Fig.~3 by van den Eijnden et al. 2018 and Fig.~2 by Russell et al. 2018).

We also stress that the non-detection of the radio counterpart would not be so surprising if the anti-correlated variability pattern between the radio and X-ray emissions (Ferrigno et al. 2014; Bogdanov et al. 2018) is ubiquitous for all tMSPs. The \nustar\ light curve shows that \src\ spent overall only $\sim30$~min in the X-ray low mode during the time span strictly simultaneous with the radio observation (this value shall be deemed however merely as a lower limit, owing to the uneven sampling of the light curve extracted from the \nustar\ datasets). Assuming that \src\ follows such an anti-correlation as well, we would expect the source to have attained a radio high mode only for $\lesssim10$\% of the whole ATCA on-source time. In Fig.~\ref{fig:lx_lradio} we also report the position occupied by \src\ on the radio:X-ray luminosity plane separately during the X-ray high and low modes (see Table~\ref{tab:statelog} and Sect.~\ref{sec:radioul}). In particular, the upper limit derived on the radio luminosity during the X-ray low mode of \src, $L_{5~{\rm GHz}}<10^{28}$~\lum, clearly stands above the region occupied by tMSPs and the NS population as a whole. Deeper radio continuum observations would be thus needed to possibly detect any radio emission from \src.

\section{Conclusions}
\label{conclusions}

In search for transitional pulsars lurking behind new unassociated \fermi/LAT gamma-ray sources, we carried out an intensive multi-wavelength observing campaign spanning the optical, X-ray and radio bands of the unidentified X-ray source CXOU\,J110926.4$-$650224 (\src). We discovered that \src\ is characterised by several hallmark phenomenological properties of tMSPs in the enigmatic sub-luminous accretion disk state. Hence, we propose it as a new tMSP candidate. 

\src\ represents the first case so far for a tMSP candidate whose GeV counterpart became detectable only after more than four years of continuous monitoring of the gamma-ray sky by the \fermi/LAT. We are currently scrutinising the LAT eight-year point source list and investigating systematically the multiband properties of several unidentified sources case by case, in an effort to select new potential gamma-ray faint tMSP candidates in this unique transitional phase. Meanwhile, we will monitor \src\ on a regular cadence to assess the evolution of its emission properties over a longer time span, and unveil possible marked changes that would signal the long-sought state transition typical of tMSPs.

\begin{acknowledgements}
We are grateful to the \xmm, \nustar\ and ATCA teams for the effort in scheduling the simultaneous ToO observations of CXOU\,J110926.4$-$650224. We acknowledge the International Space Science Institute (ISSI, Bern) that funded and hosted the international team `The disk-magnetosphere interaction around transitional millisecond pulsars'. We thank the referee for comments. FCZ acknowledges Poshak Gandhi for sharing his code to compute the prior based on the 3D model of the distribution of Galactic LMXBs. We made use of data from the ESA mission {\it Gaia} (https://www.cosmos.esa.int/gaia), processed by the {\it Gaia} Data Processing and Analysis Consortium (DPAC, https://www.cosmos.esa.int/web/gaia/dpac/consortium). Funding for the DPAC has been provided by national institutions, in particular the institutions participating in the {\it Gaia} Multilateral Agreement. We also made use of data from the Two Micron All Sky Survey, which is a joint project of the University of Massachusetts and the Infrared Processing and Analysis Center/California Institute of Technology, funded by the National Aeronautics and Space Administration (NASA) and the National Science Foundation (NSF). Some of the observations reported in this paper were obtained with the {\it Southern African Large Telescope} (SALT), as part of the Large Science Programme on transients 2016-2-LSP-001 (PI: Buckley). Some of the scientific results reported in this study are based on observations obtained with \xmm\ and \integ, which are European Space Agency (ESA) science missions with instruments and contributions directly funded by ESA Member States and NASA; \emph{The Neil Gehrels Swift Observatory}, which is a NASA/UK/ASI mission; and the \nustar\ mission, which is a project led by the California Institute of Technology, managed by the Jet Propulsion Laboratory and funded by NASA. The \emph{Australia Telescope Compact Array} is part of the Australia Telescope National Facility, which is funded by the Australian Government for operation as a National Facility managed by CSIRO. \texttt{IRAF} is distributed by the National Optical Astronomy Observatory (NOAO), which is operated by the Association of Universities for Research in Astronomy (AURA) under a cooperative agreement with the NSF. \texttt{PyRAF} is a product of the Space Telescope Science Institute, which is operated by AURA for NASA. This research has made use of the \nustar\ Data Analysis Software (NuSTARDAS) jointly developed by the ASI Science Data Center (ASDC) and the California Institute of Technology, as well as of software provided by the {\em Chandra X-ray Center} (operated for and on behalf of NASA by the Smithsonian Astrophysical Observatory [SAO] under contract NAS8--03060) in the application package \texttt{CIAO}. We made use of data supplied by the UK \swift\ Science Data Centre at the University of Leicester and of the XRT Data Analysis Software (XRTDAS). We also used softwares and tools provided by the High Energy Astrophysics Science Archive Research Center (HEASARC) Online Service, which is a service of the Astrophysics Science Division at NASA/GSFC and the High Energy Astrophysics Division of SAO. This research has made use of the Palermo BAT Catalogue and database operated at INAF - IASF Palermo.  FCZ, DFT and NR are supported by grants SGR2017-1383, AYA2017-92402-EXP and iLink 2017-1238. FCZ is also supported by a Juan de la Cierva fellowship. AP and DdM acknowledge financial support from the Italian Space Agency and National Institute for Astrophysics, ASI/INAF, under agreements ASI-INAF I/037/12/0 and ASI-INAF n.2017-14-H.0. AP also acknowledges funding from the EUs Horizon 2020 Framework Programme for Research and Innovation under the Marie Sk\l odowska-Curie Individual Fellowship grant agreement 660657-TMSP-H2020-MSCA-IF-2014. DAHB is supported by the National Research Foundation (NRF) of South Africa. JL acknowledges support from the Alexander von Humboldt Foundation. TDR is supported by a Veni grant from the Netherlands Organization for Scientific Research (NWO). MG is supported by Polish National Science Centre grant OPUS 2015/17/B/ST9/03167. Polish participation in SALT is funded by grant no. MNiSW DIR/WK/2016/07. We acknowledge support from The National Key Research and Development Program of China (2016YFA0400800), the National Natural Science Foundation of China via NSFC-11473027, NSFC-11503078, NSFC-11673013, NSFC-11733009 and the Strategic Priority Research Program `The Emergence of Cosmological Structures' of the Chinese Academy of Sciences, grant No. XDB09000000. We thank the PHAROS COST Action (CA16214) for partial support.
\end{acknowledgements}

\end{document}